\documentclass[review]{elsarticle}
\journal{Spatial Statistics}

\usepackage{lineno,hyperref}

\usepackage{natbib}

\usepackage{amsmath}
\usepackage{amsfonts}
\usepackage{graphicx}
\usepackage{url} 
\usepackage{xr-hyper}
\usepackage{lscape}
\usepackage{rotating}
\usepackage{subcaption}
\usepackage{multirow}
\usepackage{ltablex}
\usepackage{ragged2e}
\usepackage{booktabs}
\usepackage{verbatim}
\usepackage{placeins}
\usepackage{float}
\usepackage{graphicx}

\usepackage{booktabs}
\usepackage{longtable}
\usepackage{array}
\usepackage{diagbox}
\usepackage{wrapfig}
\usepackage{colortbl}
\usepackage{pdflscape}
\usepackage{tabu}
\usepackage{threeparttable}
\usepackage{threeparttablex}
\usepackage[normalem]{ulem}
\usepackage{makecell}
\usepackage{xcolor}
\usepackage{hhline,colortbl}

\usepackage{algorithm}
\usepackage{algorithmic}
\usepackage{afterpage}

\usepackage{bm}

\usepackage{tikz}

\hypersetup{
    colorlinks = true,
    citecolor = {black}, 
    urlcolor = {blue}, 
    linkcolor = {black}, 
    filecolor={black}
}

\newcommand{\appropto}{\mathrel{\vcenter{
  \offinterlineskip\halign{\hfil$##$\cr
    \propto\cr\noalign{\kern2pt}\sim\cr\noalign{\kern-2pt}}}}}

\defcitealias{KDHS2014}{KDHS, 2014}
\defcitealias{BurundiDHS:10}{BDHS, 2012}
\defcitealias{MalawiDHS:15}{MDHS, 2017}
\defcitealias{NigeriaDHS:18}{NDHS, 2019}
\defcitealias{GADM}{GADM, 2018}
\defcitealias{local2020mapping}{LBD and others, 2020}
\defcitealias{local2021mapping}{LBD and others, 2021}
\defcitealias{paige2022design}{Paige et al., 2022}
\defcitealias{dong2021space}{Dong and Wakefield, 2021}
\defcitealias{wu2021dhs}{DHS, 2021}

\defcitealias{KenyaCensus:2009}{Kenya National Bureau of Statistics, 2014}
\defcitealias{NMICS16}{Nigeria National Bureau of Statistics, 2016}
\defcitealias{MICS}{IPUMS, 2024}

\defcitealias{elev}{NOAA, 2022}

\usepackage[T1]{fontenc}
\usepackage{newpxmath}

\journal{Spatial Statistics}

\biboptions{authoryear}

\makeatletter
\DeclareRobustCommand{\textsupsub}[2]{{%
  \m@th\ensuremath{%
    ^{\mbox{\fontsize\sf@size\z@#1}}%
    _{\mbox{\fontsize\sf@size\z@#2}}%
  }%
}}
\makeatother

\begin{document}

\begin{frontmatter}

\title{Spatial fusion of DHS and MICS surveys\\with jittered and geomasked locations}


\author[NRaddress]{John Paige\corref{mycorrespondingauthor}}
\cortext[mycorrespondingauthor]{Corresponding author}
\ead{jpaige@nr.no}

\author[NTNUaddress]{Geir-Arne Fuglstad}
\author[NTNUaddress]{Andrea Riebler}

\address[NRaddress]{Norsk Regnesentral, Oslo, Norway}
\address[NTNUaddress]{Department of Mathematical Sciences, NTNU, Trondheim, Norway}

\begin{abstract}

In low- and middle-income countries, survey data is a key data source for public health monitoring. Not accounting for the anonymization applied to the GPS locations in these surveys may lead to bias, and in turn sub-optimal public health decision-making and outcomes. We propose a spatial fusion model to account both for jittering (a form of random perturbation of GPS coordinates) and geomasking (reporting only the name of the area containing the true GPS coordinates). This allows us to combine data from surveys with different kinds of positional uncertainty, reduce bias, and give more precise estimates.

We use this approach to estimate women's secondary education completion through a fusion model using both the 2018 Nigeria demographic and health survey (DHS), with jittered point locations, and the 2016 Nigeria multiple indicator cluster (MICS) survey, with geomasked locations. We show that accounting for the positional uncertainty in the surveys and accounting for both DHS and MICS surveys simultaneously can reduce bias while improving precision in parameter estimates and predictions via a simulation study and validation at cluster and areal levels. We recommend the proposed approach particularly in cases where the modeler wishes to model DHS and MICS surveys jointly, where continuously varying spatial covariates are used, and where local targeted spatial estimates are desired.

\end{abstract}


\end{frontmatter}


\section{Introduction} \label{sec:introduction}

Geostatistical modeling of health and demographic indicators at fine spatial scales is increasingly emphasized \citep{lbd2020mapping, osgood:etal:18}. This is true for a number of reasons, including the improved ability to target populations that most need assistance when aiming to reach various UN sustainable development goals (SDGs), which set targets for a number of health and demographic indicators to reach by 2030 \citep{sdgsWeb}. These SDGs explicitly state the importance of reducing inequality in health and demographic outcomes both demographically and spatially.

Since official statistics in low- and middle-income countries frequently exhibit substantial bias, third party surveys are typically considered the gold standard for data in such contexts \citep{li:etal:19,wagner:etal:18,sandefur2015political,jerven2013poor,devarajan2013africa}. Demographic and health surveys (DHS, \citealp{DHSPweb}) and multiple indicator cluster surveys \citepalias[MICS, ][]{MICS} are two kinds of surveys that are frequently used for estimating health and demographic indicators in low- and middle-income countries \citep{lbd2020mapping, osgood:etal:18, paige2022spatial, burstein:etal:18, godwin2021space,graetz:etal:18}.

Yet using DHS and MICS surveys to produce predictions at fine spatial or temporal scales can be challenging for two reasons. First, DHS and MICS surveys deliberately adjust the spatial locations of the surveyed clusters (i.e.~surveyed villages and city neighborhoods) for anonymity purposes \citep{samplingManualDHS}, resulting in uncertainty in the clusters' spatial locations known as positional uncertainty. This positional uncertainty is routinely ignored in classical and commonly used geostatistical models. Yet ignoring location anonymization in geostatistical models, has been shown to lead to attenuated parameter estimates, and can reduce estimated spatial dependency along with predictive ability \citep{cressie2003spatial, fanshawe:diggle:11, fronterre:etal:18}.

\begin{figure}
\centering
\includegraphics[width=0.55\textwidth]{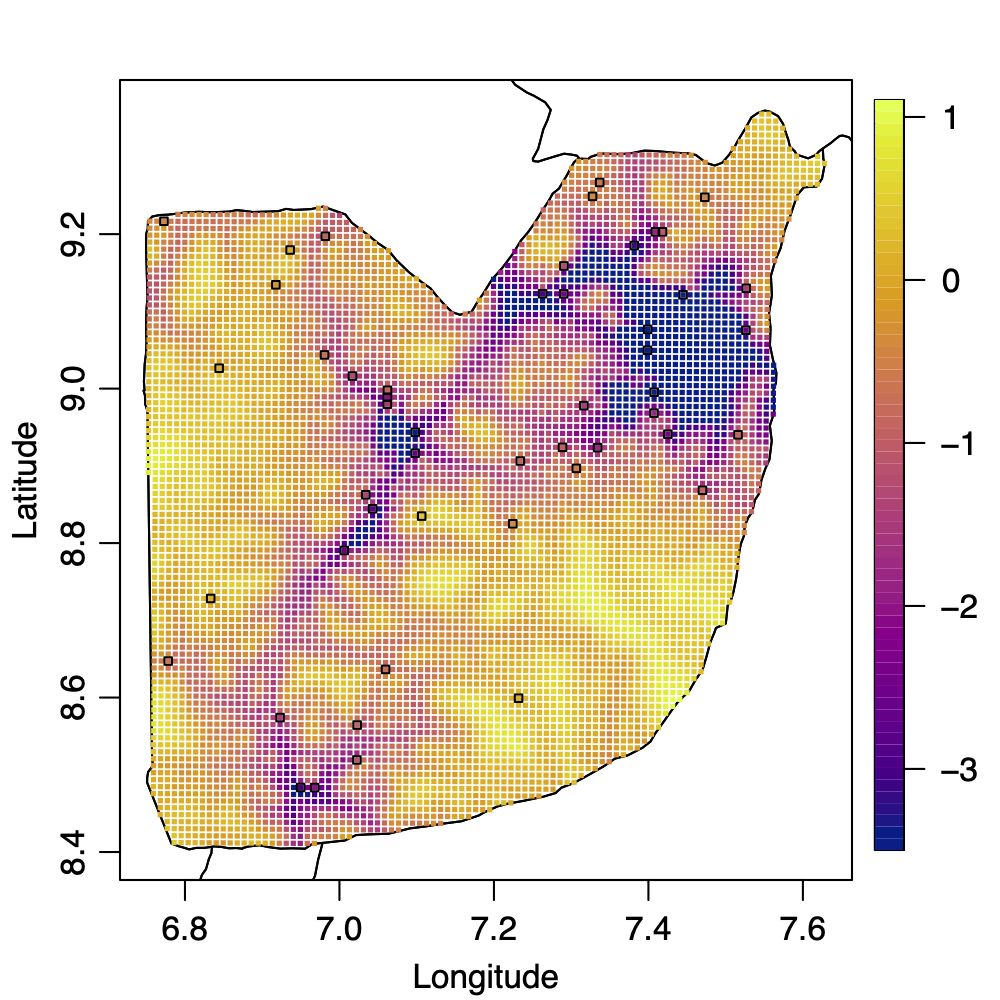} 
\caption{Illustration of positional uncertainty in FCT Abuja, Nigeria. Black points represent example locations of an observation, and colors indicate the value of a transformed covariate, healthcare inaccessibility.}
\label{fig:positionalUncertaintyIllustration}
\end{figure}

Figure \ref{fig:positionalUncertaintyIllustration} illustrates how the positional uncertainty of a single cluster might affect its response if it could be anywhere within the Federal Capital Territory (FCT) Abuja in Nigeria. Depending on the location of the cluster, the effect of the depicted covariate could vary substantially, influencing predictions for that cluster. The black points in the figure are technically integration points generated based on the numerical integration scheme described in Section \ref{sec:methods}, but can be thought of as a set of example locations that a single cluster might have.

Second, while DHS and MICS surveys can help in these contexts, they are costly and sparse \citep{paige2022design, marquez2021harmonizing}, and finer and finer scale estimation of health and demographic indicators necessitates more and more data. Data fusion modeling is then a natural approach to satisfy such demands. However, DHS and MICS surveys use two different kinds of anonymization, and hence exhibit two different kinds of positional uncertainty, making a fusion model for such data sources difficult. As a result, a fusion modeling framework has not yet been developed to jointly model responses from the surveys while accounting for their different forms of location anonymization in a principled way.

In DHS surveys, cluster locations are typically \textit{jittered}, which involves displacing them by uniformly distributed random angles and radii \citep{samplingManualDHS, DHSspatial07}. In MICS surveys, cluster locations are typically \textit{geomasked}, which means that only the administrative area containing the cluster GPS coordinates along with the cluster's urban/rural level is known \citep{khan2019multiple, NMICS16}.

While it is common to assume point locations of the modeled observations are known, some methods can account for uncertainty in the point locations in certain contexts. For example, \citet{cressie2003spatial} propose a way to perform kriging in the presence of positional uncertainty by adjusting the covariates and the covariance function, using a pseudolikelihood approach for parameter estimation. Their approach assumes a linear model, however, and a Berkson type error model \citep{carroll:etal:06} for the locations, which is less suitable than a classical error model for this context since the true locations that influence outcomes are fixed, and observed locations are assigned based on the truth. \citet{fanshawe:diggle:11} propose a general method for accounting for positional uncertainty using a combination of Monte Carlo likelihood maximization for inference and numerical integration for prediction. They note that likelihood maximization took 72 hours for modeling data at 148 spatial locations, however. \citet{wilson2021estimation} proposed using integrated nested Laplace approximations within Markov chain Monte Carlo (MCMC, \citealp{gomez:rue:18}) to handle general forms of point location anonymization and the resulting positional uncertainty within Bayesian hierarchical models with a spatial random effect, but they note that the method took 52 hours to generate 1000 MCMC samples.

More recently, \citet{altay2022fast} proposed a numerical integration method for fitting spatial Bayesian hierarchical models under jittering. They showed that the method could be used to speed up inference considerably compared to \citet{fanshawe:diggle:11} and \citet{wilson2021estimation}. \citet{altay2024impact} further demonstrated that the model could be used effectively in the presence of raster based covariates, and that jittering could, if not handled properly, attenuate raster based covariate effect size estimates. \citet{bakar2026bayesian} propose a Bayesian model for DHS data that accounts for jittering, but they do not consider geomasking or joint modeling of DHS and MICS surveys, and the method as written can only make predictions at observed clusters.

In this work, we extend the method proposed in \citet{altay2022fast} to account for more general forms of positional anonymization with a focus on geomasking. This poses a greater challenge than jittered data in DHS surveys in that there is substantially more positional uncertainty. Our proposed method is able to jointly and accurately model DHS and MICS data, and it can do so at a fraction of the time of existing approaches that can be applied in such contexts, while still accounting for the positional uncertainty present in DHS and MICS data under a consistent and principled statistical framework.

In Section \ref{sec:data} we discuss the details of the data we use in our analysis, including the DHS and MICS surveys. In Section \ref{sec:methods} we introduce a classical geostatistical model for health and demographic indicators, and show how it can be modified to create a joint spatial model accounting for anonymization of spatial positions in DHS and MICS surveys jointly. Section \ref{sec:motivatingSim} illustrates how accounting for positional uncertainty and using the proposed fusion model can improve inference and predictions via a simulation study. In Section \ref{sec:application}, we apply some of the considered models to the chosen application, estimating women's secondary education prevalence in Nigeria. In Section \ref{sec:validation} we perform a validation against sampled clusters and direct areal estimates of the proposed model along with several others for comparison. Finally, we summarize our conclusions in Section \ref{sec:conclusions}.

\section{Data} \label{sec:data}

\subsection{Household survey data}

Both the 2018 Nigeria demographic and health survey (NDHS) \citep{NigeriaDHS:18} and the 2016 Nigeria multiple indicator cluster survey (NMICS) \citepalias{NMICS16} sample enumeration areas (EAs), villages and city neighborhoods defined by the 2006 Population Census. EAs are defined by the census, and are officially defined as either urban or rural. The NDHS selects a fixed number of EAs from the urban and rural parts of each of the 37 Admin1 areas (36 states and the federal capital territory), the administrative districts below the national level. The EAs are selected with probability proportional to size (PPS) sampling, where number of households is used as a proxy for population size. In the 2018 NMICS, a fixed number of EAs are selected from the FCT, 35 Admin1 areas, and the three senatorial districts in each of the two remaining states: Kano and Lagos. Hence, the NDHS is stratified with strata equal to the urban/rural parts of each area, making 74 strata, whereas the NMICS strata are the 35 Admin1 areas and 6 senatorial districts, making 41 strata in total. The NDHS samples 25 households from each selected EA to be included in the survey, whereas the NMICS samples 16 households from each selected EA. Both use simple random sampling for sampling households within selected EAs to form the sampled clusters. In total, the NDHS and NMICS have 1,379 and 2,195 clusters and 13,980 and 11,037 women aged 20--29 respectively.

Response rates for both surveys were very high, with women's response rates being 94\% for the NMICS and 99\% for the NDHS. We consider women without any formal education to have not completed their secondary education, in part because literacy rates are lower among women without a formal education than women who have only attended primary school.

For privacy reasons both the DHS and MICS anonymize the cluster centroid point locations in two different ways. In DHS surveys, it is common to \textit{jitter} cluster locations, which involves displacing them by uniformly distributed random angles and radii \citep{samplingManualDHS, DHSspatial07}. NDHS cluster centroids in urban areas are jittered by up to 2$\, \mathrm{km}$, and 99\% of rural locations are jittered by up to $5\, \mathrm{km}$ with 1\% of them being jittered by up to $10\, \mathrm{km}$. In MICS surveys, cluster locations are \textit{geomasked}, which means that only the administrative area containing the cluster GPS coordinates along with the clusters urban/rural level is known \citep{khan2019multiple, NMICS16}. NMICS clusters are geomasked to the  level of the areal stratification variable. The locations of the NMICS clusters therefore typically have much higher positional uncertainty than the NDHS cluster locations. Such levels of positional uncertainty make it difficult for fine scale geostatistical models to incorporate MICS data.

\begin{figure}
\centering
\includegraphics[width=.49 \textwidth]{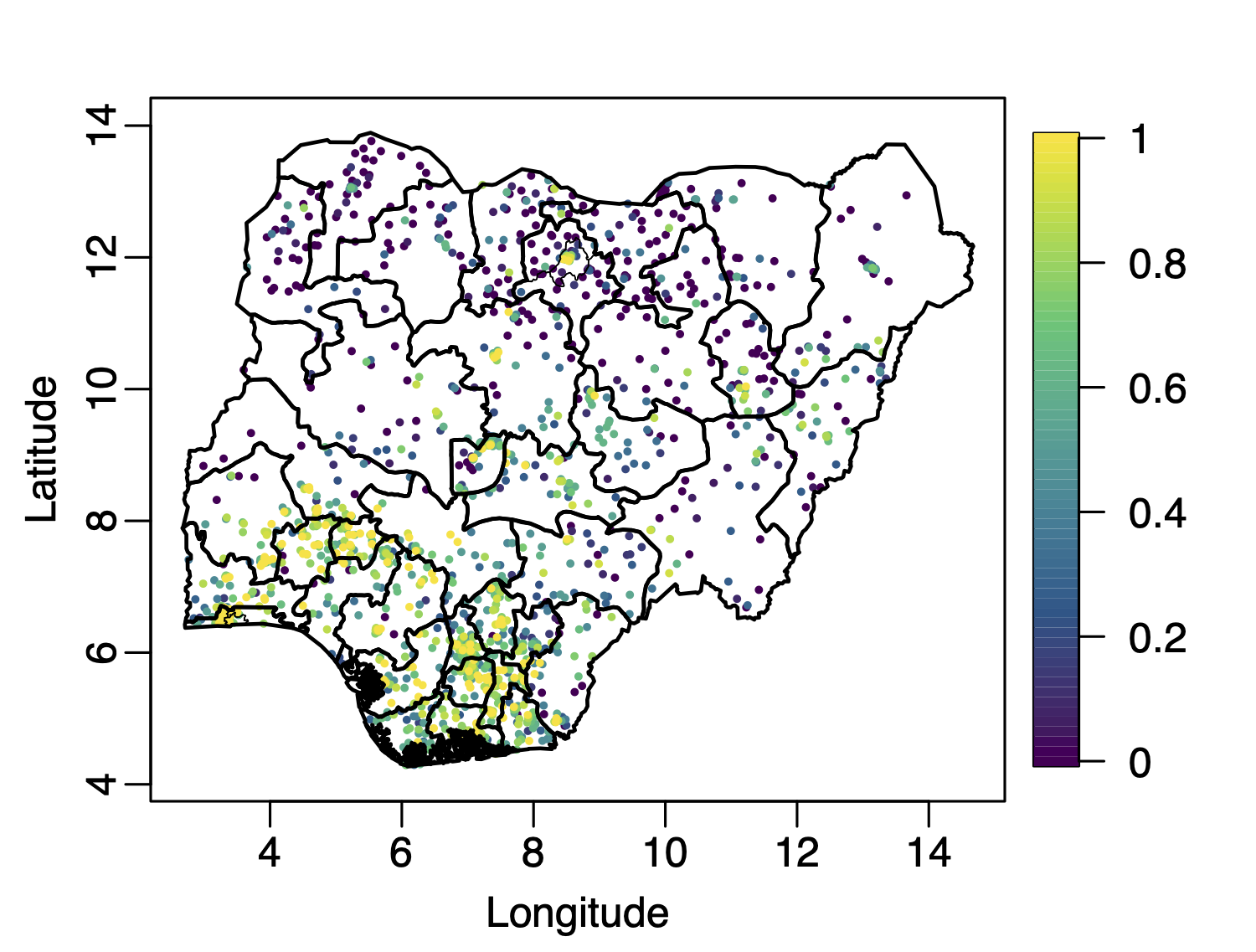} \includegraphics[width=.49 \textwidth]{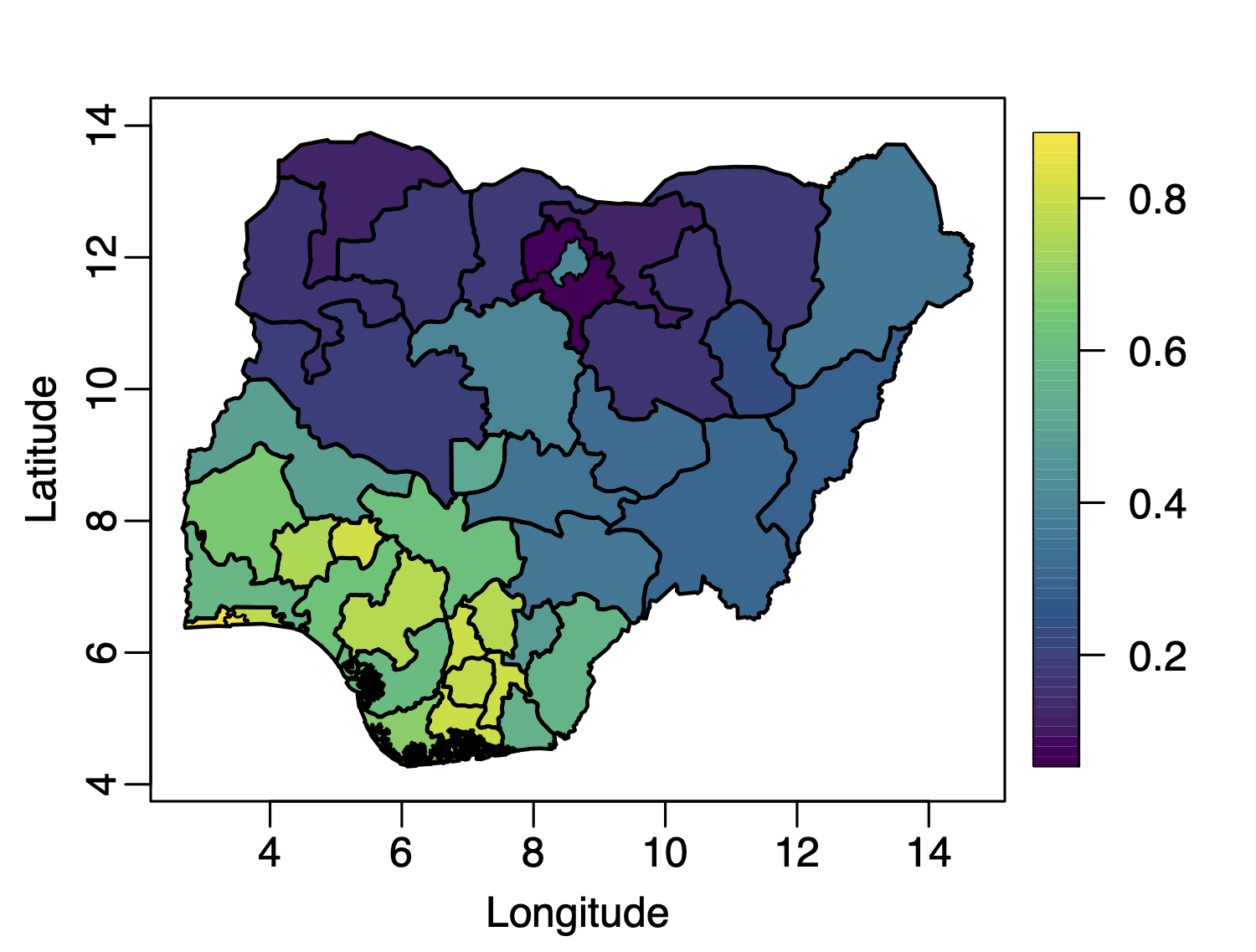}
\caption{Raw, unweighted prevalences of women's secondary education from DHS (left) and MICS (right) surveys. Admin1 areas and 6 senatorial districts are outlined in thick and thin lines respectively.}
\label{fig:data}
\end{figure}

The raw, unweighted responses of both surveys are plotted spatially in Figure \ref{fig:data}, where individual responses are aggregated within clusters for the DHS data and within geomasked MICS strata for the MICS data. The figure indicates women's secondary education completion prevalence is generally highest in the southern and southwestern parts of Nigeria, and lowest in the north and northwestern parts of the country, with medium rates near 0.5 in the central, central-eastern, and northeastern parts of the country, although spatial correlation and survey based corrections are not accounted for in the figure.

\subsection{Covariate rasters} \label{sec:covariates}

As in \citet{altay2024impact}, we consider five different raster-based covariates: population density as estimated by WorldPop \citep{pop}, urban/rural classifications that we generate based on population density, access to healthcare as measured in travel time \citep{weiss2018global}, elevation \citepalias{elev}, and distance to rivers and lakes \citep{riverLake}. We transform most of these covariates and then normalize them, subtracting the mean and dividing by the standard deviation. The transformation and renormalization of the covariate rasters is intended to improve the interpretability of the relative effect sizes and the accuracy of the numerical integration. After calibrating population density within each admin2 area to be proportional to the estimated total urban and rural populations, the resulting population density raster is transformed via the $\log(x + 1)$ function. The healthcare inaccessibility variable is also $\log(x+1)$-transformed, whereas elevation is square root transformed, and distance to rivers and lakes is not transformed prior to normalization.

Unlike the other covariates, the urbanicity classification raster is generated based on population density using the methods in \citet{paige2022design} and \citet{paige2022spatial} in order to carefully calibrate aggregation from urban/rural pixels to the Admin1 areal level. Specifically, urbanicity is derived by thresholding the untransformed population density surface so as to ensure urban and rural populations are as close to the estimated numbers as possible, with any deviations removed through calibration. It is neither transformed nor normalized, since it is a binary variable.

\begin{figure}
\centering
\subcaptionbox{Population density}[.49\textwidth]{\includegraphics[width=\linewidth]{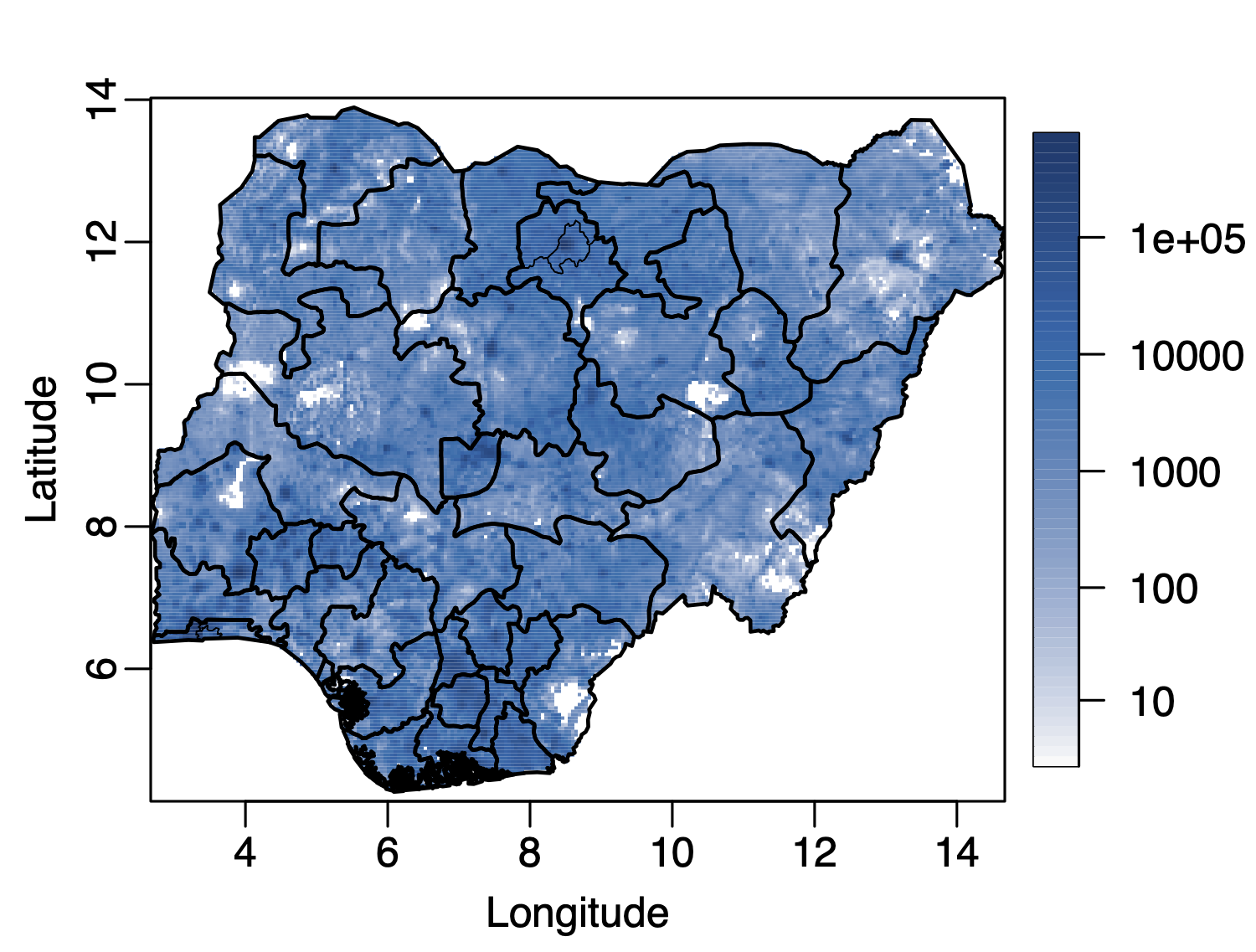}} \subcaptionbox{Urbanicity}[.49\textwidth]{\includegraphics[width=\linewidth]{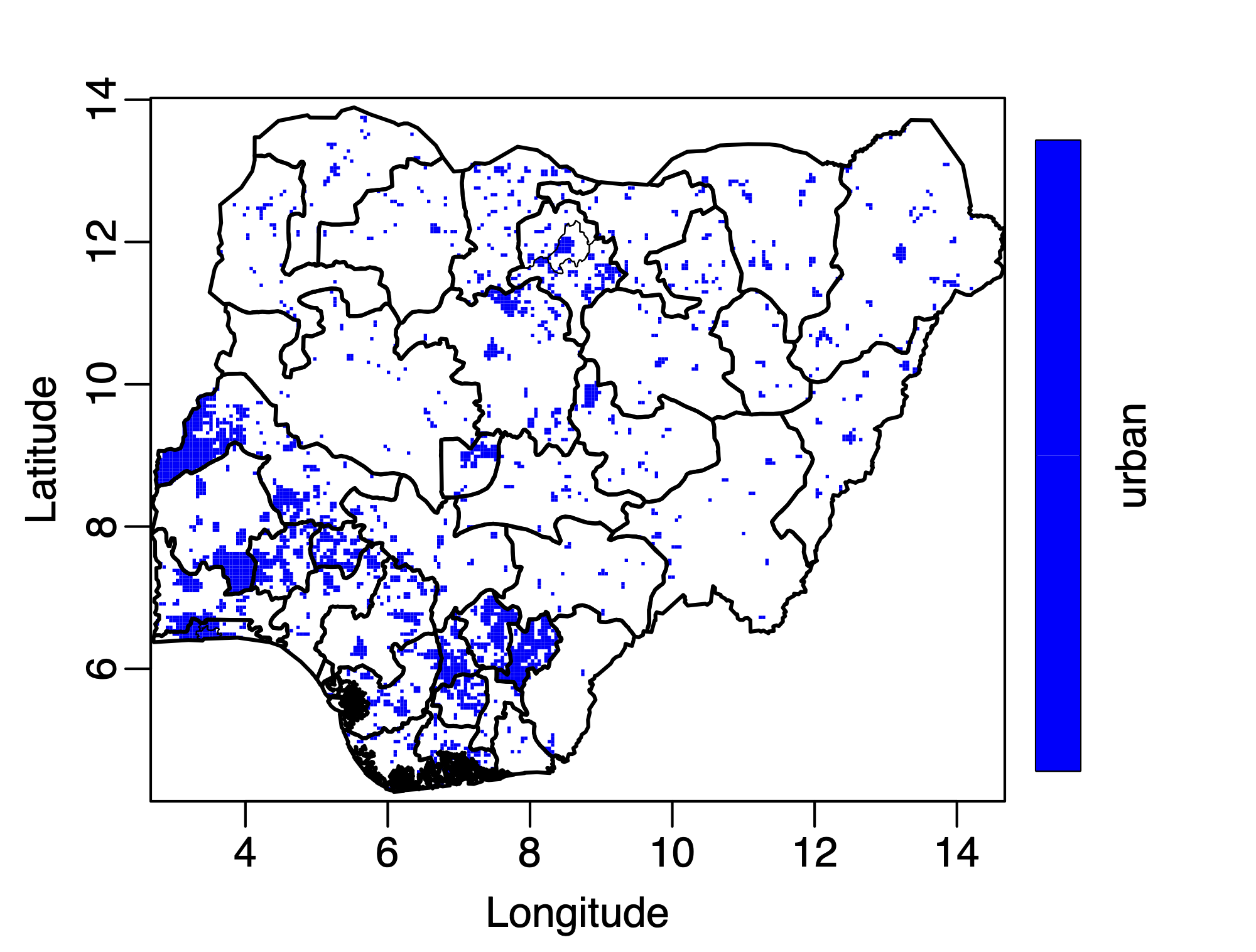}} \\
\subcaptionbox{Healthcare inaccessibility}[.49\textwidth]{\includegraphics[width=\linewidth]{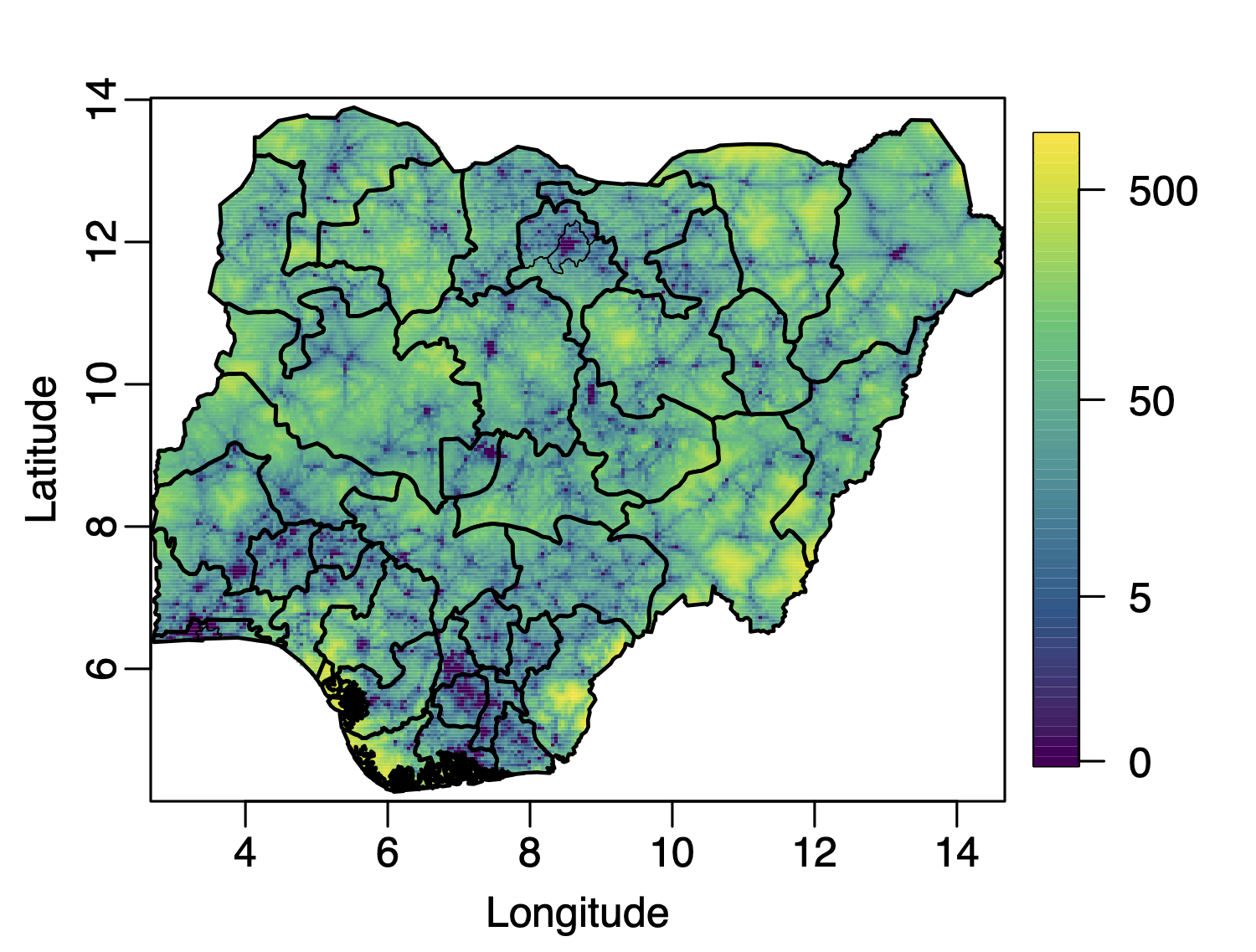}} \subcaptionbox{Elevation (m)}[.49\textwidth]{\includegraphics[width=\linewidth]{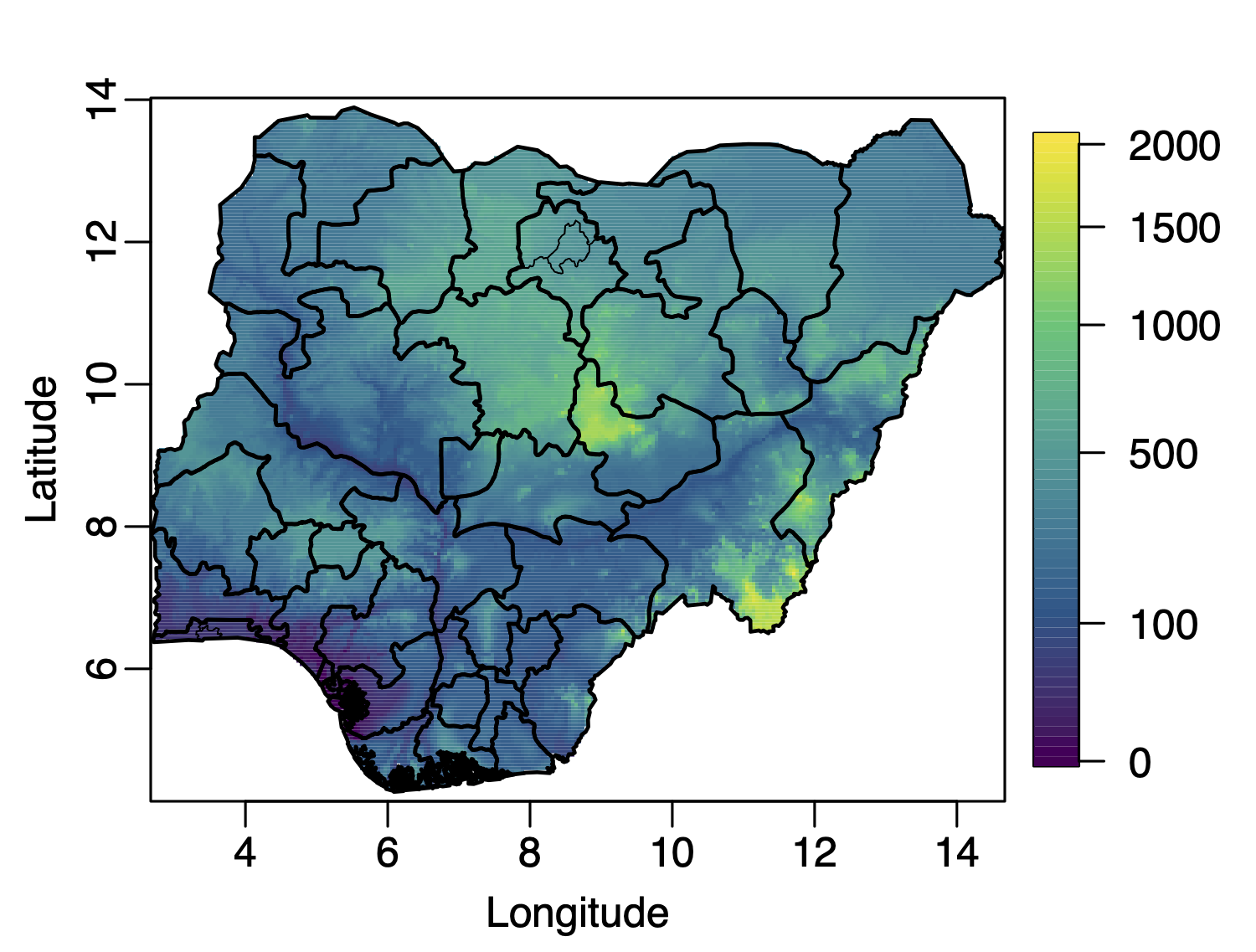}} \\
\subcaptionbox{Distance to rivers, lakes}[.49\textwidth]{\includegraphics[width=\linewidth]{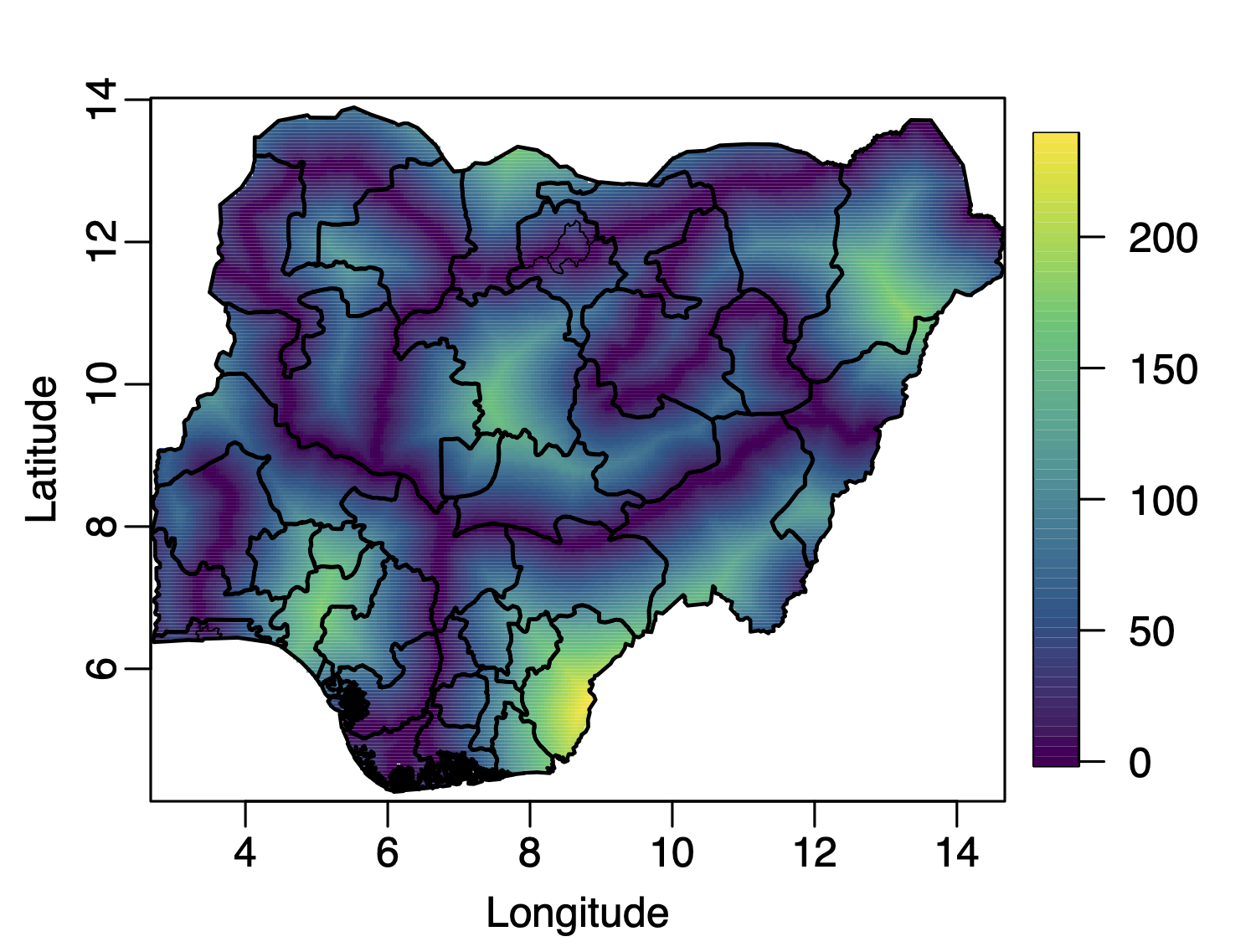}}
\caption{Covariates included in the analysis prior to transformation. Admin1 areas and 6 senatorial districts are outlined in thick and thin lines respectively. Color scales are transformed using the transformations associated with each covariate.}
\label{fig:covariates}
\end{figure}

\section{Methods} \label{sec:methods}

\subsection{Spatial risk model}

We let $r(t)$ be the typical spatial probability (risk) of a woman aged 20--29 completing her secondary education at location $t \in \mathcal{D}$, with $\mathcal{D} \subset \mathbb{R}^2$ representing the spatial domain, Nigeria, with $m$ modeled areas. This risk is then modeled as:
\begin{equation}
\mbox{logit}(r(t)) = \boldsymbol{d}(t)^T \boldsymbol{\beta} + u(t). \label{eq:spatialRiskModel}
\end{equation}
The intercept and spatial fixed effects are given by $\boldsymbol{d}(t)^T \boldsymbol{\beta}$. We model the spatial random effects, $u(t)$, using a reparameterization of the Besag-York-Molli\'{e} (BYM) model referred to as BYM2 \citep{riebler:etal:16}, and it can be represented as,
$$ u(t) = \frac{1}{\sqrt{\tau}} \left( \sqrt{\phi} w_{A[t]} + \sqrt{1 - \phi} v_{A[t]} \right), $$
where $\boldsymbol{w} = (w_1, \ldots, w_m)^T$ represents spatially structured random effects, and $\boldsymbol{v}$ represents iid unstructured random effects. Both $\boldsymbol{w}$ and $\boldsymbol{v} = (v_1, \ldots, v_m)^T$ have mean zero and unit generalized variance \citep{sorbye201439}, and $A[t]$ gives the index of the modeled area associated with location $t$.  The generalized precision of $u$ is given by $\tau$, and $\phi$ is the proportion of variance that is structured.

We assume that $A[t]$ is the MICS stratum associated with $t$ so that the BYM2 effect varies at the level of the geomasked area (hence, it varies at the Admin1 level except in Lagos and Kano, where it is modeled at the senatorial district level).

\subsection{Response model}

We model the responses for a given survey, whether it be DHS or MICS. Let $y_c$, $c=1,\ldots, C$, be the number of women aged 20--29 who completed their secondary education at cluster $c$ for that survey, which contains $C$ clusters. We assume $t_c \in \mathcal{D}$ for $c=1,\ldots, C$ are the true locations of the clusters in the survey. The response model is then:
\begin{align}
y_c \mid r^{\tiny \mbox{clust}}_c &\sim \text{Bin}(n_c, \ r^{\tiny \mbox{clust}}_c), \nonumber \\
\mbox{logit}(r^{\tiny \mbox{clust}}_c) &= \mbox{logit}(r(t_c)) + \epsilon_c \nonumber \\
&= \boldsymbol{d}(t_c)^T \boldsymbol{\beta} + u(t_c) + \epsilon_c, & c=1,\ldots, C. \label{eq:responseModel}
\end{align}
Here, $r^{\tiny \mbox{clust}}(t_c)$ is the risk for women specifically in cluster $c$, which is the same as $r(t_c)$ except it contains an additional cluster level random intercept, $\epsilon_c$, representing the influence of EA/cluster specific conditions on the logit risk. The $\epsilon_c$ for $c=1,\ldots,C$ are independent and identically distributed (iid) zero mean Gaussian noise with variance given by $\sigma_\epsilon^2$. In addition, $n_c$ is the number of women aged 20--29 in cluster $c$ that responded to the survey.

We assume the spatial risk model \eqref{eq:spatialRiskModel} as well as the response model \eqref{eq:responseModel} apply to both the DHS and MICS surveys. Conditional on the risk, the responses are assumed to be independent across clusters as well as surveys.

Throughout this work we use penalized complexity (PC) priors \citep{simpson:etal:17}, which allow for a general framework for penalizing random effects, shrinking models towards a base model without random effects or with simplified structure in available random effects. The penalization is on a natural scale based on the Kullback-Leibler distance from the base model. We place a PC prior on the BYM2 precision and nugget standard deviation parameters so that $P(1/\sqrt{\tau} < 1) = P(\sigma_\epsilon < 1) = 0.1$, and a PC prior on the BYM2 proportion of structured variance, $\phi$, so that $P(\phi < 0.5) = 2/3$ as recommended in \citet{riebler:etal:16}.

\subsection{Accounting for positional uncertainty in DHS and MICS surveys}

Throughout this section, we assume that a cluster $c$ in a given survey has true location $t_c \in \mathcal{D}$ for $c \in 1, \ldots, C$, and observed location $s_c \in \mathcal{D}$ in the case of DHS data or $s_c \subset \mathcal{D}$ in the case of MICS data. Note that $s_c$ is a point location for DHS data and an observed geomasked area (stratum) for MICS data.

\subsubsection{Likelihoods under positional uncertainty}
We extend the model proposed in \citet{altay2022fast} and \citet{altay2024impact} by generalizing it to account for geomasking in a joint model for DHS and MICS surveys. We will write the likelihood conditional on the underlying risk as:
\begin{align}
    \pi \big(y_c, s_c | r(\cdot) \big) &= \int_{\mathcal{D}} \pi \big(y_c, s_c | r(t_c) \big) \, \pi(t_c) \ \mathrm{d}t_c, \notag \\
    &= \int_{\mathcal{D}} \pi \big(y_c | r(t_c) \big) \, \pi(s_c | t_c) \, \pi(t_c) \ \mathrm{d}t_c, \label{eq:posErrMod}
\end{align}
for $c = 1, \ldots, C$. Again, we assume the cluster level responses are conditionally independent given the underlying risk, both within and between surveys.

We assume that, \textit{a priori}, the distribution of true cluster locations is proportional to population density, since the EAs forming the clusters are constructed in the census to have roughly equal populations. Hence, for both DHS and MICS surveys, we assume that $\pi(t_c) \propto q(t_c)$ for each cluster $c = 1, \ldots, C$ in a given survey, where $q(\cdot)$ is the population density as a function of spatial location as calculated from the WorldPop data described in Section \ref{sec:covariates}.

Note that while the official urbanicity level of all clusters in both the DHS and MICS surveys is known, the official urbanicity designation at any given spatial location is unknown. We must therefore classify urbanicity spatially ourselves as discussed in Section \ref{sec:covariates}, and as demonstrated in \citet{paige2022design}. In order to ensure consistency between inference and spatial prediction, we constrain the support of the prior over true cluster locations, $\pi(t_c)$, to have the correct urbanicity level for MICS clusters, i.e.~to have support $\{t \in A[t_c] : \mbox{urbanicity}(t) = \mbox{urb}^{\tiny \mbox{MICS}}[c]\}$, where $\mbox{urbanicity}(t)$ is the urbanicity classification raster generated as described in Section \ref{sec:covariates}, and where $A[t]$ is the MICS stratum containing cluster location $t \in \mathcal{D}$. We opt against restricting the support of the priors for the DHS cluster locations based on urbanicity, since the DHS cluster locations are known to a much higher degree of accuracy, and also are sometimes not close enough to the nearest pixel classified with a matching urbanicity level. In both cases, the urbanicity covariate used in the response model for a given cluster is its true known urbanicity, while the classified urbanicity raster is used in pixel level and aggregated predictions.

\subsubsection{DHS: Positional uncertainty distribution and computational details}

The DHS cluster locations are jittered within a relatively short distance of the respective true locations. Given the maximum jittering distance, DHS cluster locations are jittered in a uniformly distributed random direction from 0 to $2\pi$, and by a uniformly distributed radius from 0 to the maximum jittering distance. The jittering distribution of the observed urban DHS location $s^{\tiny \mbox{DHS}}_c$ is:
$$ \pi(s^{\tiny \mbox{DHS}}_c \mid t^{\tiny \mbox{DHS}}_c) \propto \frac{\mathbb{I}\{A_2[s^{\tiny \mbox{DHS}}_c] = A_2[t^{\tiny \mbox{DHS}}_c]\} \cdot \mathbb{I}\{ |s^{\tiny \mbox{DHS}}_c - t^{\tiny \mbox{DHS}}_c|<2 \}}{ |s^{\tiny \mbox{DHS}}_c - t^{\tiny \mbox{DHS}}_c|}, $$ 
where coordinates are easting/northing in kilometers, $\mathbb{I}\{\cdot\}$ is an indicator function. Here, $A_2[t]$ is the Admin2 area associated with location $t$. Note that, generally, DHS cluster locations are never jittered outside of Admin1 boundaries for surveys before 2008, or Admin2 boundaries for DHS surveys after 2008 \citep{perez2013guidelines}. Hence, for DHS surveys taken before 2008, the $A_2[t]$ term should be changed from an Admin2 to an Admin1 level term. 

Since 99\% of rural clusters are jittered up to 5$\, \mathrm{km}$, with 1\% of rural clusters jittered up to 10$\, \mathrm{km}$, rural DHS cluster locations have the following jittering distribution:
\begin{align*}
\pi(s^{\tiny \mbox{DHS}}_c \mid t^{\tiny \mbox{DHS}}_c) \propto& \frac{\mathbb{I}\{A_2[s^{\tiny \mbox{DHS}}_c] =A_2[t^{\tiny \mbox{DHS}}_c]\}}{|s^{\tiny \mbox{DHS}}_c - t^{\tiny \mbox{DHS}}_c|} \cdot \\
&\bigg[\frac{99 \cdot \mathbb{I}\{|s^{\tiny \mbox{DHS}}_c - t^{\tiny \mbox{DHS}}_c|<5\}}{100} + \frac{\mathbb{I}\{|s^{\tiny \mbox{DHS}}_c - t^{\tiny \mbox{DHS}}_c|<10\}}{100}\bigg]. 
\end{align*}

Following \citet{altay2022fast} and \citet{altay2024impact}, for each cluster $c$ in the given survey, we evaluate \eqref{eq:posErrMod} numerically via,
\begin{equation}
    \pi \big(y_c, s_c | r(\cdot) \big) \approx
    \sum_{k = 1}^{K}
    \alpha_{ck} \, \pi \big(y_c | r(\tilde{t}_{ck}) \big) \, \pi \big(s_c | \tilde{t}_{ck} \big) \, \pi \big(\tilde{t}_{ck} \big),
    \label{eq:intSchemeDHS}
\end{equation}
for integration weights $\alpha_{ck}$ and associated with integration points $\tilde{t}_{ck}$, for $k=1, \ldots, K$, where integration points and weights are as given in \citet{altay2022fast} for DHS clusters. For both the DHS and MICS integration schemes, we take $\omega_{ck} \propto \alpha_{ck} \pi \big(s_c | \tilde{t}_{ck} \big) \pi \big(\tilde{t}_{ck} \big)$, and require $\sum_{k} \omega_{ck}=1$ for each fixed cluster $c$. We can renormalize in this way provided the renormalization constants are functions of fixed quantities, and in this case they are only a function of quantities fixed prior to the analysis: population density, the covariate rasters, and the number of integration points. The resulting numerical integral for the likelihood is then calculated as:
\begin{equation}
\pi \big(y_c, s_c | r(\cdot) \big) \stackrel{\cdot}{\propto}
    \sum_{k = 1}^{K}
    \omega_{ck} \, \pi \big(y_c | r(\tilde{t}_{ck}) \big),
\label{eq:intSchemeRenorm}
\end{equation} 
where $\stackrel{\cdot}{\propto}$ denotes `approximately proportional to', and the constant of proportionality depends only on quantities fixed prior to the analysis.

To produce areal predictions, we marginalize over the cluster specific risk effects at the pixel level, forming the `smooth risk' estimator from \citet{paige2022spatial}. Pixel level predictions are then aggregated to the areal level, weighting by population density to produce areal averages.

\subsubsection{MICS: Positional uncertainty distribution and computational details}

While our broader framework for handling geomasked MICS data is similar to our framework for the jittered DHS data, the higher positional uncertainty of the geomasked MICS data required a more careful numerical integration scheme over the possible cluster locations, as naive approaches can lead to numerical instability, slow computation, unidentifiability, and poor inference. To eliminate these effects, we introduce a marginalization scheme over the cluster specific risk effects via a Gauss-Hermite quadrature rule, as well as an optimization scheme automating the careful selection of integration points for the cluster locations themselves that incorporates information about the spatial covariates and strata.

Unlike jittered DHS data, MICS cluster point locations are known only up to the MICS stratum they lie in and their urban/rural level. Hence, the distribution for the MICS cluster positions given the true locations is much simpler, with probability mass function $\pi(s^{\tiny \mbox{MICS}}_c \mid t^{\tiny \mbox{MICS}}_c) = \mathbb{I}\{s^{\tiny \mbox{MICS}}_c \ni t^{\tiny \mbox{MICS}}_c\}$, where `$\ni$' denotes `contains', and recalling that $s^{\tiny \mbox{MICS}}_c$ is the geomasked area for cluster $c$ of the MICS survey.

Naively approximating a posterior over the joint distribution of $\boldsymbol{\beta}$, $u(\cdot)$, and cluster specific risk effects via a Laplace approximation is numerically unstable when accounting for the levels of positional uncertainty common in MICS data. This is due in particular to confounding between the cluster specific risk effects and the positional uncertainty. Instead, we introduce a marginalization scheme over the cluster specific risk effects via a Gauss-Hermite (GH) quadrature rule as
\begin{equation}
\pi \big(y_c | r(\tilde{t}_{ck}) \big) = \int_{-\infty}^\infty \pi \big(y_c | r(\tilde{t}_{ck}), \epsilon_c \big) \pi(\epsilon_c) \ \mathrm{d}\epsilon_c \approx \sum_{l = 1}^{L} v_l \, \pi \big(y_c | r(\tilde{t}_{ck}), \epsilon^{(l)} \big).
\label{eq:GH}
\end{equation}
This vastly reduces the optimization space (since there is one cluster specific risk effect per cluster we marginalize rather than estimate), improves numerical stability, and allows for more accurate inference in the presence of the high positional uncertainty inherent to geomasked MICS data.

 We use the \verb|fastGHQuad| R package \citep{fastGHQuad} to obtain integration points $\epsilon^{(1)}, \ldots, \epsilon^{(L)}$ and weights $v_1, \ldots, v_L$, with $L=10$ integration points. A Laplace approximation is formed over the remaining effects. While we found this integration scheme was far more numerically stable, and was generally sufficient, in some cases the Hessians at the posterior optimum were non-positive definite. In such cases, we increase the number of GH integration points by 10, and rerun the optimization, repeating this process until a positive-definite Hessian is obtained. We find the resulting integration scheme is often more robust than the naive Laplace approximation performed jointly over smooth spatial and cluster specific risks used in \citet{altay2022fast} and \citet{altay2024impact}, and $L$ never needed to be increased beyond 30 in this work.

In addition to integrating out cluster specific risk effects, we choose positional integration points in a very different way for MICS clusters than for DHS clusters. Since covariates vary over the support of the more diffuse distributions of cluster positions, we choose integration points that can better represent the full range of possible combinations of covariate values. Since the integration points and weights are calculated only once, it is possible to choose them carefully without spending much computational effort overall. More information on choosing the integration points and weights, along with illustrations, is given in the Supplement in Section \ref{sec:supplementMICSintegration}. We choose $K$ to be 100 for MICS clusters, 16 for urban DHS clusters, and 21 for rural DHS clusters in order to achieve a good blend of computational speed and accuracy.

\subsubsection{The IHME fusion model for MICS and DHS data} \label{sec:ihme}

\citet{lbd2020mapping} uses a spatial fusion method that does not account for positional uncertainty in DHS data, and handles the positional uncertainty in MICS data differently, with code available at \url{https://github.com/ihmeuw/lbd}. Rather than attempting to directly integrate the likelihood for each MICS cluster over the uncertainty in the cluster's true location as in \eqref{eq:posErrMod}, \citet{lbd2020mapping} uses a weighted pseudolikelihood approach similar to in \eqref{eq:intSchemeRenorm}, except where equivalent population density-based weights are used to average log probabilities rather than probabilities:
$$ \log \pi \big(y_c, s_c | r(\cdot) \big) \approx
    \sum_{k = 1}^{K}
    \omega_{ck} \, \log \pi \big(y_c | r(\tilde{t}_{ck}) \big). $$
While taking the weighted average on a log scale may seem innocuous, Jensen's inequality implies the resulting estimator is not equivalent, and can lead to bias in the resulting likelihood estimate for each cluster that does not disappear as the number of integration points (or clusters) increases.

Rather than a fully Bayesian approach, \citet{lbd2020mapping} uses empirical Bayes, conditioning on posterior estimates of the hyperparameters in the predictive distribution. In addition, \citet{lbd2020mapping} also uses a different method for choosing integration points associated with each weight. First, a number of random points are drawn, where the total is about one point per populated km$^2$ per MICS stratum with probability proportional to population density. Then $k$-means is used to cluster the random points into about one cluster per 1,000km$^2$ of populated area, and the cluster centroids are used as the integration points. This corresponds to 930 integration points for the whole country, and does not stratify integration point selection by urban/rural level. For reference, our method uses 8,100 MICS integration points. The weights of each integration point are then proportional to the number of random points in the corresponding $k$-means cluster. Unlike the proposed method of choosing integration points, this method does not directly penalize clusters for having variability in their spatial covariates.

We use essentially the same spatial risk and response models as the other approaches, but with the likelihood for MICS clusters replaced by the pseudolikelihood described above, and without accounting for positional uncertainty in the DHS data. Aside from differences in the likelihood, there is also a difference in the response model in that we consider two implementations, one using a BYM2 spatial effect for comparability with the other approaches, and one using a stochastic partial differential equation (SPDE) spatial effect \citep{Lindgren:etal:11}. The SPDE effect version is used in \citet{lbd2020mapping}, while the BYM2 effect version is included for posterity.

To produce areal predictions, cluster/EA specific risk effects are included at the 5 km$^2$ pixel level when sampling from the posterior predictive distribution. Since there is not necessarily one EA per pixel, this can lead to a mismatch in the uncertainty between the areal predictive distribution and the underlying target population \citep{paige2022spatial}. For reference, there are 35,579 pixels with nonzero population in the domain, far fewer than the 662,529 EAs from which clusters are sampled. We would therefore expect pixel level inclusion of EA specific risk effects to inflate areal level uncertainties, although nonlinearities in the way such uncertainties can be added and averaged makes the effect difficult to predict.

\subsubsection{Considered modeling approaches for DHS and MICS data}

We consider 6 approaches denoted by M\textsubscript{d}, M\textsubscript{D}, M\textsubscript{M}, M\textsubscript{\tiny DM}, IHME\textsubscript{\tiny SPDE}, and IHME\textsubscript{\tiny BYM2} depending on whether the MICS data is included, whether positional uncertainty in the DHS and MICS datasets is fully accounted for, and whether the approach uses IHME's framework and SPDE latent spatial effect model or not:\\[.3cm]
\noindent
\textbf{M\textsubscript{\tiny d}:} Only DHS data is included, and the observed locations are assumed to be the true locations.\\[.3cm]
\noindent
\textbf{M\textsubscript{\tiny D}:} Only DHS data is included, and jittering is accounted for using the proposed method.\\[.3cm]
\noindent
\textbf{M\textsubscript{\tiny M}:} Only MICS data is included, and geomasking is accounted for using the proposed method.\\[.3cm]
\noindent
\textbf{M\textsubscript{\tiny DM}:} Both DHS and MICS data is included, and positional uncertainty is accounted for in both datasets using the proposed method.\\[.3cm]
\noindent
\textbf{IHME\textsubscript{\tiny SPDE}:} An existing approach in the literature based on \citet{lbd2020mapping}. Both DHS and MICS data is included, and positional error is handled as described in Section \ref{sec:ihme}. Uses a stochastic partial differential equation (SPDE) spatial effect.\\[.3cm]
\textbf{IHME\textsubscript{\tiny BYM2}:} Like the IHME\textsubscript{\tiny SPDE} approach, but uses a BYM2 spatial effect.\\[.3cm]

While not an existing method in the literature, we also include a BYM2 version of the IHME model in the Supplement for comparison. The IHME approaches use the original source code involving integrated nested Laplace approximations (INLA, \citealt{rue2009approximate}), while the other models are implemented using the template model builder (TMB, \citealt{kristensen:etal:15}) package, since their likelihoods involve a mixture on a probability scale, which is not supported by INLA. While the M\textsubscript{\tiny d} model does not involve a mixture, we implement it in TMB for consistency with the proposed models. See \citet{osgood2023statistical} for a detailed comparison of INLA and TMB for spatial modeling.

\section{Motivating simulation study: Modeling geomasked MICS data} \label{sec:motivatingSim}

In order to assess the predictive and inferential performance of existing methods for accounting for MICS data, in comparison to the proposed approach for handling MICS data, we run an initial simulation study using realistic parameters based on the considered surveys. We also use this simulation study to assess the feasibility of modeling MICS data without any supplemental DHS data at both Admin1 and Admin2 levels.

For the spatial risk model in \eqref{eq:spatialRiskModel}, the marginal variance $\sigma^2$ and cluster variance $\sigma_\epsilon^2$, along with the intercept and covariate effects, are all set to the values estimated for the proposed model fitted to the data (model M\textsubscript{\tiny DM} in Table \ref{tab:parSummary1}). However, rather than simulating directly from the proposed model with $u$ being modeled with a BYM2 effect, we simulate from a `neutral' model, where $u$ is simulated via the SPDE continuously-indexed spatial modeling approach of \citet{Lindgren:etal:11} so as not to bias the results of the simulation study towards the proposed model. Although this means that the marginal and cluster variance parameters used in the simulation model will correspond to something slightly different than the fitted values in Table \ref{tab:parSummary1}, the parameters will still be realistic since they represent similar quantities. We also set the effective range of $u$ to be 200km, roughly a fifth of the spatial domain diameter.

In total, we simulate 100 spatial risk surfaces along with 100 MICS and DHS surveys from the empirical distribution of population sizes per cluster and fixing the number of clusters per urban/rural part of each stratum to that of the true MICS and DHS surveys. True locations, target population, risks, and burden in each cluster are simulated jointly using the SUMMER package \citep{Li:2024aa}, where cluster locations are chosen with probability proportional to population density in each survey stratum.

\subsection{Evaluation metrics}

The considered approaches are compared based on their areal prevalence predictions at the MICS stratum level. We compare each of these to the simulated target population prevalence in terms of a number of scoring rules. Specifically, we consider mean square error (MSE), continuous ranked probability score (CRPS), interval score (IS), fuzzy empirical coverage, and fuzzy CI width. Note that CRPS and IS are strictly proper and proper scoring rules respectively, so that the expected value of the scores of true model is strictly or nonstrictly smaller than those of incorrect models \citep{gneiting:raftery:07}. Fuzzy empirical coverage and fuzzy CI width \citep{paige2022bayesian} are employed to eliminate artifacts due to deterministic discrete confidence/credible intervals so that, for 80\% fuzzy CIs, the expected fuzzy empirical coverage should be close to 80\% regardless of the ability of a deterministic CI to capture exactly 80\% of a predictive distribution's mass. They are based on fuzzy intervals as defined in \citet{geyer2005fuzzy}.

Scoring rules and metrics are evaluated for a given target $y$ and predictive distribution $P$, where $y$ is a prevalence and is therefore on a probability scale:
\begin{align*}
\mbox{MSE}(y; P) &= (y - E_{P}[y])^2 \\
\mbox{CRPS}(y; P) &= \int_{-\infty}^\infty (P(y \leq \tilde{y}) - \mathbb{I}\{y \leq \tilde{y}\})^2 \ \mathrm{d} \tilde{y} \\
\mbox{IS}_\alpha(y; P) &= (u - l) + \frac{2}{\alpha}(y - u) I \left\{y > u \right\} + \frac{2}{\alpha}(l - y) I \left\{y < l \right\} \\
\mbox{Cvg}_\alpha^{\tiny \mbox{fuzzy}}(y; P) &= E_{P}[\mathbb{I}\left\{\mbox{L}^{\tiny \mbox{fuzzy}}_\alpha \leq y \leq \mbox{U}^{\tiny \mbox{fuzzy}}_\alpha \right \}] \\
\mbox{Width}_\alpha^{\tiny \mbox{fuzzy}}(y; P) &= E_{P}[\mbox{U}^{\tiny \mbox{fuzzy}}_\alpha - \mbox{L}^{\tiny \mbox{fuzzy}}_\alpha],
\end{align*}
where $[l, u]$ is a credible interval for $y/n$ based on predictive distribution $P$ with significance at most $\alpha$, where $l = \sup\{x : P(x) \leq \alpha/2\}$ and $u = \inf\{x : P(u) < 1- \alpha/2\}$. Similarly, $[\mbox{L}^{\tiny \mbox{fuzzy}}_\alpha, \ \mbox{U}^{\tiny \mbox{fuzzy}}_\alpha]$ is a fuzzy credible interval for $y$ based on predictive distribution $P$ with significance equal to $\alpha$. Note that $l$ and $u$ in the expression for $\mbox{IS}_\alpha(y; P)$ are the fixed lower and upper endpoints for the $\alpha$ significance equal tailed credible interval given by predictive distribution $P$. Fuzzy endpoints $\mbox{L}^{\tiny \mbox{fuzzy}}_\alpha$ and $\mbox{U}^{\tiny \mbox{fuzzy}}_\alpha$ are not used in the interval score, since they would not result in a proper scoring rule.

While the data generating mechanisms used parameters with fixed, known values, the spatial effect was inherently different from the considered approaches used for predictions. Hence, the `best' parameters for fitted models from a predictive standpoint are not necessarily those used in the true data generating process, particularly for the population simulated using the SPDE spatial effect. Hence, scores of the parameter estimates against the true data generating parameters reflect the ability of the approaches to infer the true data generating parameters rather than the best parameters for predictive purposes.

\subsection{Simulation study results}

\begin{table}[ht]
\centering
\resizebox{1\linewidth}{!}{
\begin{tabular}{rrrrrrr}
  \hline
Model & Bias & MSE & CRPS & 95\% IS & 95\% Cvg & 95\% Width \\
  \hline
M\textsubscript{\tiny d} & 0.005 & 0.001 & 0.022 & 0.178 & 0.93 & 0.143 \\[-4pt]
M\textsubscript{\tiny D} & -0.001 & 0.001 & 0.022 & 0.195 & 0.93 & 0.139 \\[-4pt]
M\textsubscript{\tiny M} & -0.002 & 0.001 & 0.020 & 0.165 & 0.93 & 0.128 \\[-4pt]
M\textsubscript{\tiny DM} & -0.002 & 0.001 & 0.016 & 0.132 & 0.92 & 0.099 \\[-4pt]
IHME\textsubscript{\tiny SPDE} & 0.003 & 0.002 & 0.025 & 0.296 & 0.76 & 0.102 \\[-4pt]
IHME\textsubscript{\tiny BYM2} & 0.002 & 0.001 & 0.016 & 0.138 & 0.97 & 0.126 \\
   \hline
\end{tabular}
}
\caption{SPDE simulation study prediction Admin1 level scores and metrics.}
\label{tab:sim1}
\end{table}


Table \ref{tab:sim1} shows the main simulation study results for the populations simulated with SPDE spatial effects, and Table \ref{tab:sim1BYM2} in the Supplement contains results from the BYM2-simulated populations for reference.

While DHS and MICS datasets can be individually useful, using both simultaneously in a principled joint model seems to improve the CRPS of population and parameter estimates as evidenced by the performance of the M\textsubscript{\tiny DM} model. Over the 100 simulated populations, the average CRPS for the M\textsubscript{\tiny M} and M\textsubscript{\tiny D} models were 0.020 and 0.022 respectively, yet the CRPS of the joint M\textsubscript{\tiny DM} model was tied with the IHME\textsubscript{\tiny BYM2} model for the lowest at 0.016. The IHME\textsubscript{\tiny SPDE} model, however, had the worst areal prediction CRPS with 0.025. With the exception of the IHME\textsubscript{\tiny SPDE} model, the credible intervals of all approaches were satisfactory, being within 3\% of the nominal level, and the estimator bias was also very small for all models being under 5 people per thousand in magnitude for all models. Interestingly, the IHME\textsubscript{\tiny BYM2} approach achieved CRPS tied for the best despite, like the IHME\textsubscript{\tiny SPDE} model, having estimates of the urban and population density effects that were biased by between 0.43 and 0.45 in magnitude, positive for urban and negative for population density. Covariate effect estimate scores are given in Tables \ref{tab:sim1par} and \ref{tab:sim1parBYM2} in the Supplement for the SPDE and BYM2 simulated populations respectively.

The results also indicate that accounting for positional uncertainty in a principled way can improve both parameter and population estimates by reducing attenuation bias in parameter estimates. While the M\textsubscript{\tiny D} and M\textsubscript{\tiny d} approaches had similar areal prediction CRPS of 0.022, the M\textsubscript{\tiny D} and noticeably reduced bias in the population density estimate of -0.021 compared to -0.152 for the M\textsubscript{\tiny d} approach, where the large bias of the M\textsubscript{\tiny d} approach is due to attenuation bias. The M\textsubscript{DM} approach had the best CRPS for the urban and population density effects of 0.061 and 0.085 respectively due to using both DHS and MICS datasets while accounting for positional uncertainty in both. The IHME models' CRPS for the urban and population density effects were substantially worse, above 0.4. The IHME\textsubscript{\tiny SPDE} model's poor population and parameter estimate performance is particularly notable since the data generating process used an SPDE rather than BYM2 spatial effect. All approaches were able to identify that the healthcare inaccessibility, elevation, and distance to rivers and lakes covariates are spurious, estimating near zero effect sizes for each of them.

While the M\textsubscript{\tiny M} approach had better CRPS for its areal predictions than the M\textsubscript{\tiny D} approach due to there being more MICS than DHS data, the M\textsubscript{\tiny D} approach had better CRPS for covariate effect estimates. This is due to the fact that the MICS data has more positional uncertainty than the DHS data. However, the  M\textsubscript{\tiny DM} model had the best CRPS for covariate effect estimates, since it uses both datasets while accounting for positional uncertainty in both.

Overall, we found that the proposed M\textsubscript{\tiny DM} had the best performance in terms of CRPS and interval score for the main parameter effects (urban and population density effects) as well as Admin1 level predictions, the one exception being that the M\textsubscript{\tiny D} had a better but still comparable interval score for the effect of population density at 0.813 compared to 0.871.

\section{Application to women's secondary education} \label{sec:application}

\begin{sidewaysfigure}
\centering
\includegraphics[width=1.1 \textwidth]{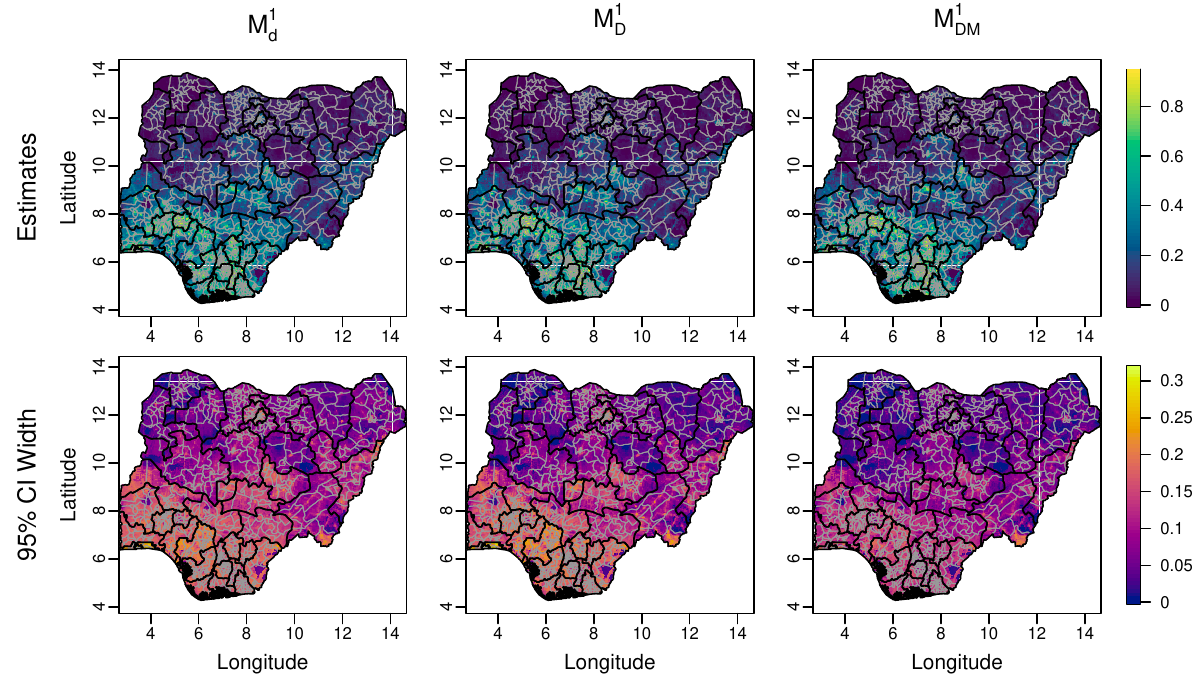}
\caption{5km pixel level women's secondary education prevalence predictions (top row) and 95\% credible interval widths (bottom row). Admin2 area boundaries are given in light grey while MICS stratum boundaries are given in black.}
\label{fig:edPixel1}
\end{sidewaysfigure}

\begin{sidewaysfigure}
\centering
\includegraphics[width=1.1 \textwidth]{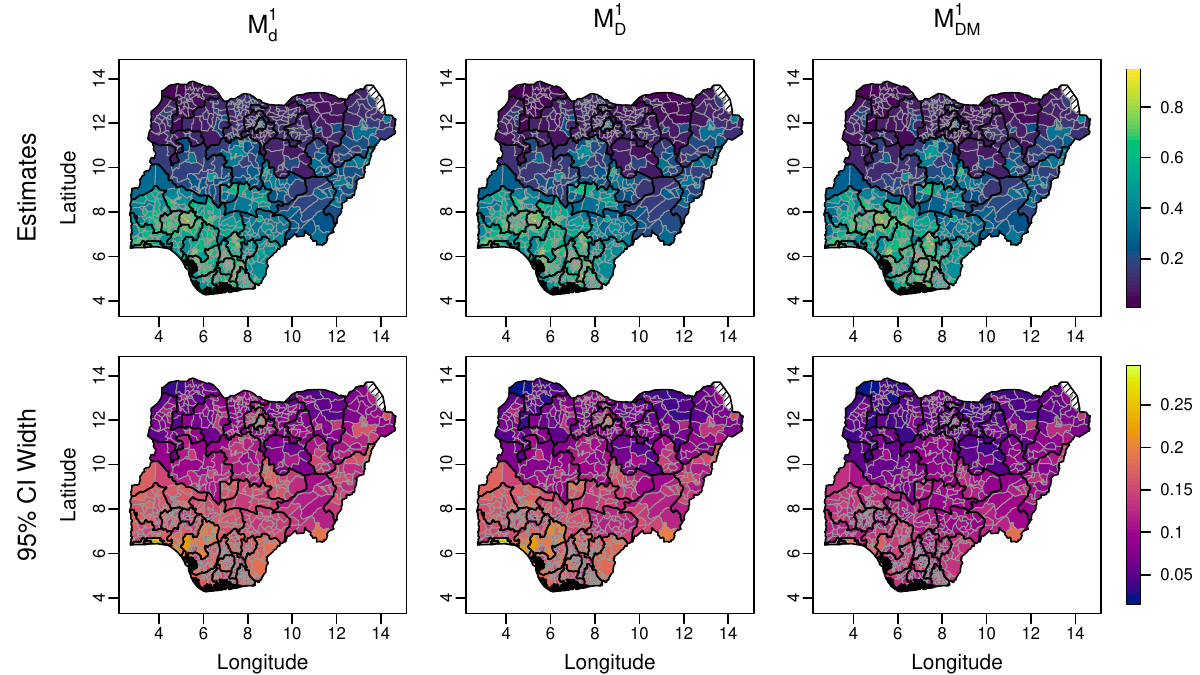}
\caption{Admin2 level women's secondary education prevalence predictions (top row) and 95\% credible interval widths (bottom row). Admin2 area boundaries are given in light grey while MICS stratum boundaries are given in black.}
\label{fig:edAdmin21}
\end{sidewaysfigure}

\begin{sidewaysfigure}
\centering
\includegraphics[width=1.1 \textwidth]{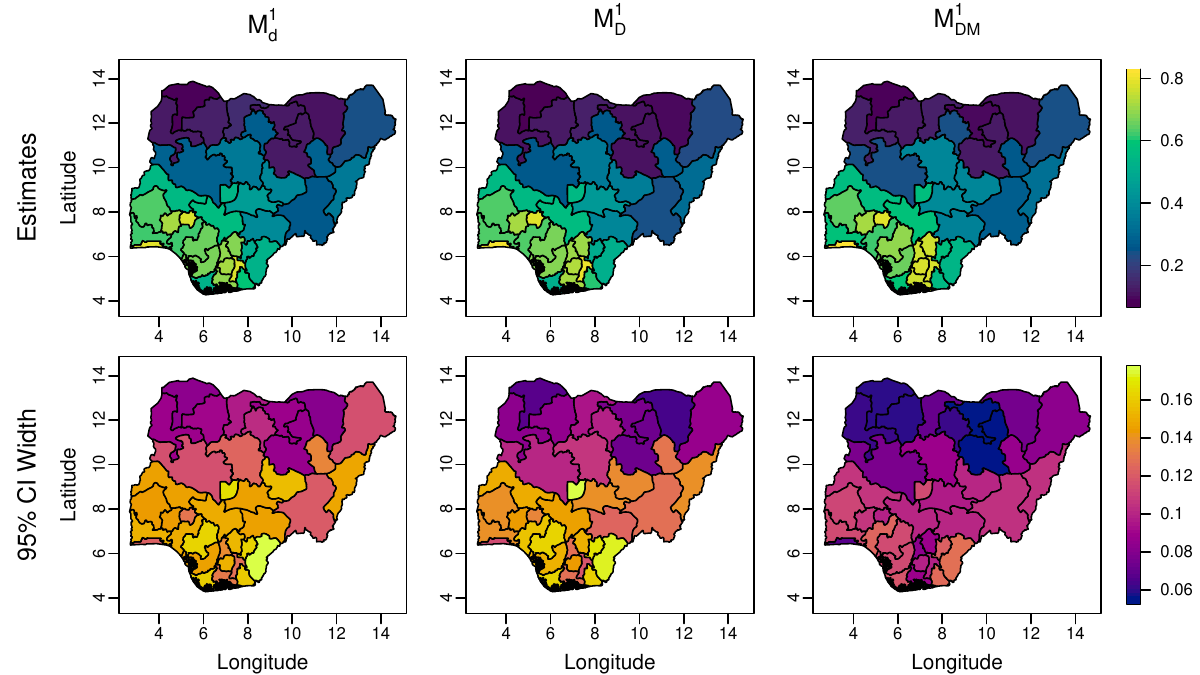}
\caption{Admin1 level women's secondary education prevalence predictions (top row) and 95\% credible interval widths (bottom row).}
\label{fig:edAdmin11}
\end{sidewaysfigure}

We apply the approaches to women's secondary education in Nigeria using both the 2018 NDHS and 2016 NMICS. We focus on the M\textsubscript{\tiny d}, M\textsubscript{\tiny D}, and M\textsubscript{\tiny DM} approaches, since they show how accounting for positional uncertainty in DHS data, and how adding in MICS data jointly with DHS data M\textsubscript{\tiny DM} affects results. In addition, we find the M\textsubscript{\tiny DM} approach performs the best in terms of CRPS in Section \ref{sec:validation} via cluster and Admin1 level validation. Posterior means along with 95\% CI widths are plotted spatially at the 5km pixel, Admin2, and Admin1 levels in Figures \ref{fig:edPixel1}-\ref{fig:edAdmin11}. Parameter posterior summary statistics for these approaches are given in Table \ref{tab:parSummary1}.

Differences in parameter estimates are most pronounced between the M\textsubscript{\tiny DM} approach and the others. The M\textsubscript{\tiny d} and M\textsubscript{\tiny D} approaches have fairly similar parameter estimates with moderate differences in the estimates (and 95\% CI widths) of the urban effect, 0.31 (0.50) and 0.33 (0.51) respectively, and cluster variance, 0.65 (0.67) and 0.64 (0.65) respectively, and a slight difference in the estimate of the effect of population density, 0.83 (0.41) and 0.99 (0.49) respectively. The M\textsubscript{\tiny DM} approach had a similar urban effect estimate, but a higher estimate of the effect of population density, with a posterior mean (and 95\% CI width) of 1.10 (0.45). This estimate is closer to the M\textsubscript{\tiny D} than the M\textsubscript{\tiny d} approach, reflecting the positional uncertainty in the data as well as the inclusion of the MICS survey.

Figures \ref{fig:edPixel1}-\ref{fig:edAdmin11} show that the M\textsubscript{\tiny DM} approach has the lowest posterior average uncertainty at pixel and areal levels due to its ability to borrow strength across both surveys. Differences between the M\textsubscript{\tiny d} and M\textsubscript{\tiny D} approaches are also noticeable in the figures, albeit less pronounced.

\begin{table}[ht]
\centering
\resizebox{1.1\linewidth}{!}{
\begin{tabular}{@{}r@{\hspace{0.5em}}r@{\hspace{0.25em}}c@{\hspace{0.5em}}r@{\hspace{0.25em}}c@{\hspace{0.5em}}r@{\hspace{0.25em}}c@{\hspace{0.5em}}r@{\hspace{0.25em}}c@{\hspace{0.5em}}r@{\hspace{0.25em}}c@{}}
& \multicolumn{2}{c}{\textit{\textbf{M\textsubscript{\tiny DM}}}} & \multicolumn{2}{c}{\textit{\textbf{M\textsubscript{\tiny M}}}} & \multicolumn{2}{c}{\textit{\textbf{M\textsubscript{\tiny D}}}} & \multicolumn{2}{c}{\textit{\textbf{M\textsubscript{\tiny d}}}} & \multicolumn{2}{c}{\textit{\textbf{IHME\textsubscript{\tiny SPDE}}}} \\
& \textbf{Est} & \textbf{95\% CI} & \textbf{Est} & \textbf{95\% CI} & \textbf{Est} & \textbf{95\% CI} & \textbf{Est} & \textbf{95\% CI} & \textbf{Est} & \textbf{95\% CI} \\
\toprule
\textbf{(Int)} & -2.15 & (-2.34, -1.96) & -2.36 & (-2.89, -1.82) & -2.07 & (-2.28, -1.84) & -1.71 & (-1.94, -1.36) & -1.84 & (-3.50, -0.19) \\
\textbf{Urban} & 0.35 & (0.15, 0.56) & 0.42 & (-0.16, 0.97) & 0.33 & (0.08, 0.59) & 0.31 & (0.06, 0.56) & 1.72 & (1.59, 1.86) \\
\textbf{Health.~Inacc.} & -0.21 & (-0.34, -0.10) & -0.31 & (-0.67, 0.06) & -0.22 & (-0.34, -0.10) & -0.20 & (-0.30, -0.09) & -0.04 & (-0.10, 0.02) \\
\textbf{Elevation} & 0.18 & (0.05, 0.31) & 0.18 & (-0.17, 0.55) & 0.17 & (0.03, 0.31) & 0.14 & (0.02, 0.26) & 0.13 & (0.05, 0.21) \\
\textbf{Dist.~Riv.~Lakes} & 0.11 & (-0.01, 0.24) & 0.15 & (-0.17, 0.46) & 0.11 & (-0.02, 0.23) & 0.09 & (-0.03, 0.22) & 0.02 & (-0.06, 0.10) \\
\textbf{Population} & 1.10 & (0.87, 1.32) & 1.20 & (0.50, 1.86) & 0.99 & (0.76, 1.25) & 0.83 & (0.63, 1.04) & 0.16 & (0.05, 0.27) \\
$\boldsymbol{\sigma^2}$ & 0.59 & (0.35, 0.95) & 0.72 & (0.41, 1.16) & 0.64 & (0.37, 1.02) & 0.65 & (0.37, 1.04) & 0.84 & \multicolumn{1}{c}{--} \\
$\boldsymbol{\phi}$ & 0.89 & (0.52, 0.99) & 0.87 & (0.53, 0.99) & 0.87 & (0.47, 0.99) & 0.86 & (0.47, 0.99) & \multicolumn{1}{c}{--} & \multicolumn{1}{c}{--} \\
\textbf{Range (km)} & \multicolumn{2}{c}{--} & \multicolumn{2}{c}{--} & \multicolumn{2}{c}{--} & \multicolumn{2}{c}{--} & 1220 & \multicolumn{1}{c}{--} \\
$\boldsymbol{\sigma_{\epsilon}^2}$ & 1.63 & (1.45, 1.83) & 1.60 & (1.07, 2.21) & 1.47 & (1.25, 1.70) & 1.43 & (1.23, 1.64) & 1.89 & \multicolumn{1}{c}{--} \\
\midrule
\textbf{Time (min.)} & \multicolumn{2}{c}{7.24} & \multicolumn{2}{c}{4.93} & \multicolumn{2}{c}{2.71} & \multicolumn{2}{c}{0.45} & \multicolumn{2}{c}{0.91} \\
\bottomrule
\end{tabular}
}
\caption{Posterior parameter central estimates and 95\% credible intervals, and computation time for model fitting, parameter estimation, and prediction.}
\label{tab:parSummary1}
\end{table}

\FloatBarrier

\section{Model validation} \label{sec:validation}

\subsection{Validation overview} \label{sec:validationOverview}
In order to evaluate the considered approaches, we perform two kinds of validation: validation of Admin1 level areal predictions, and validation of left out clusters. We wished to validate both at the cluster and areal levels in part to simply explore a new possible way to validate individual cluster level predictions with unknown spatial locations, which has potential applications in validating aggregation models \citep{paige2022spatial}, and in part due to the fact that areal and cluster level prediction may ultimately have different needs. We discuss this in greater detail later in this section, and in Section \ref{sec:conclusions}.

When validating against left out individual clusters, we perform a stratified 20-fold validation, with 10 folds consisting entirely of DHS data, and 10 consisting of MICS data. For DHS survey data, validation folds are stratified by Admin1 areas and urbanicity, whereas for the MICS survey, validation folds are stratified by the 41 MICS stratum areas and urbanicity. We ensure that the number of clusters in any one stratum differs by at most 1 across all 10 folds for a given survey. Since the IHME and M\textsubscript{\tiny DM} approaches depend on data from both surveys, they must each be fit 20 times to generate 20 sets of predictions each. Since the M\textsubscript{\tiny d} and M\textsubscript{\tiny D} approaches only use the DHS data, we fit them and generate predictions 11 times---10 times for each fold of the DHS data left out, and one additional time using all of the DHS data, since leaving out different MICS folds results in the same model under those two approaches.

In the case where we validate areal predictions, we leave out 10 of the above 20 folds in a single area at a time---five (randomly selected) folds from each survey. We then produce Admin1 level survey-weighted direct estimators using the \verb|svyglm| function from the \verb|survey| package \citep{Lumley:2023}, with a Horvitz-Thompson type variance estimate \citep{horvitz:thompson:52} based on the left out DHS data, the left out MICS data, and a precision-weighted average of the two. Here, \verb|svyglm| produces an estimator by solving the classical set of survey-weighted score equations for fitting generalized linear models to survey data (see, e.g.~\citealt{binder:83, lumley:scott:17}). Finally, we compare each approach's Admin1 level predictions against the direct estimate as recommended in \citepalias{wu2021dhs}.

Note that since the direct estimate, unlike the model based estimate, includes sampling variability, it is necessary to add such sampling variability into the model based estimate so that the prediction represents a prediction for the left out data rather than for the entire population in the area. To do so, we follow the \citet{gascoigne2023estimating} approach, and calculate final validation scores and metrics on a probability scale.

Letting $Y_i$ be the left out direct estimate for the prevalence of women's secondary education completion in area $i = 1,\ldots,37$, and $\hat{Y}_i$ be the model based estimate, we consider the difference $\hat{Y}_i - Y_i$. Generally, the closer this is to zero the better the model is at producing areal estimates. Since $\hat{Y}_i$ and $Y_i$ are produced from different data sources, they are independent estimators. This means that if one were to sample $M$ independent replicates from their sampling distributions, $Y_i^{(1)}, \ \ldots, \ Y_i^{(M)}$ from the asymptotic distribution of $Y_i$, and $\hat{Y}_i^{(1)}, \ \ldots, \ \hat{Y}_i^{(M)}$ from the posterior distribution of $\hat{Y}_i$, then $\hat{Y}_i^{(1)} - Y_i^{(1)}, \ \ldots, \ \hat{Y}_i^{(M)} - Y_i^{(M)}$, would be $M$ samples from the distribution of $\hat{Y}_i - Y_i$. Using the evaluation metrics in the section below, we can then compare the distribution of $\hat{Y}_i - Y_i$ against 0, the target value. This validation scheme has the advantage that it targets predictions for a population as opposed to targeting spatial predictions. Also, it accounts for the survey weights.

\subsection{Aggregating evaluation metrics}
We first consider cluster level validation. Due to positional uncertainty, we are only able to validate cluster level predictions in the case where the cluster location is unknown. The considered validation task is then to predict the response for a given out of sample cluster conditional on the observed (with positional uncertainty) location.

Letting $\mathcal{O}_1^j, \ldots, \mathcal{O}_{10}^j$ be the sets of indices of the out of sample clusters for the 10 cross validation folds for survey $j \in \{\mbox{DHS}, \mbox{MICS}\}$, and let $\mathcal{I}_1^j, \ldots, \mathcal{I}_{10}^j$ be the in sample data as well as other information used for prediction such as the observed locations of the out of sample clusters. Given a score or metric $S(y; P)$ for a single cluster prevalence $y$ and predictive distribution $P$, we consider the cluster level validation score averaging over each fold $k$ and survey $j$ as, 
$$ S_{\tiny \mbox{clust}}^j = \frac{1}{N^j} \sum_{k=1}^{10} \sum_{c \in \mathcal{O}_k^j} n_c^j \cdot S\left(y_c^j; \, P\big(\cdot \mid \mathcal{I}_k^j \big)\right), $$
where $P$ is the predictive distribution under a given approach for a given set of sample cluster locations and target population sizes.  Here, $y_c^j$ and $n_c^j$ are the observed prevalence and number of sampled individuals in cluster $c$ and survey $j$, and $N^j$ are the number of individuals in survey $j$. We also take a weighted average of the scores for each survey, 
$$S_{\tiny \mbox{clust}}^{\tiny \mbox{DHS+MICS}} = \frac{N^{\tiny \mbox{DHS}}}{N^{\tiny \mbox{DHS}} + N^{\tiny \mbox{MICS}}} S_{\tiny \mbox{clust}}^{\tiny \mbox{DHS}} + \frac{N^{\tiny \mbox{MICS}}}{N^{\tiny \mbox{DHS}} + N^{\tiny \mbox{MICS}}} S_{\tiny \mbox{clust}}^{\tiny \mbox{MICS}},$$
as a measure of overall performance across both surveys, where the weights are proportional to the sample sizes.

In the case of areal validation, let $\hat{Y}_i$ be the areal prediction for Admin1 area $i$, $i=1,\ldots,37$, and let $Y_i^j$ be the direct estimate for area $i$ and survey $j$, where $j=\mbox{DHS}, \mbox{MICS}, \mbox{DHS+MICS}$. Here, $j=\mbox{DHS+MICS}$ corresponds to the case where model based areal predictions are compared against the precision weighted average of the DHS and MICS direct estimates. We calculate a simple, unweighted average of metrics across areas:
$$ S_{\tiny \mbox{areal}}^j = \frac{1}{37} \sum_{i=1}^{37} S\left(0; \, P_{\hat{Y}_i-Y_i^j}\right), $$
where $P_{\hat{Y}_i-Y_i^j}$ is the distribution of $\hat{Y}_i-Y_i^j$ calculated as described in Section \ref{sec:validationOverview}. Note that we can compare the same model based estimator, $\hat{Y}_i$, against three different direct estimators, which gives an idea of how well the model based estimator can predict data from each survey separately as well as altogether.

\subsection{Validation results} \label{sec:validationResults}

\begin{table}[ht]
\centering
\begin{tabular}{rrrrrr}
\toprule
\hspace{1em}\textbf{Approach} & \textbf{CRPS} & \textbf{80\% IS} & \textbf{80\% Cvg} & \textbf{80\% Width} & \textbf{Time (min.)} \\
\bottomrule
\addlinespace[0.3em]
\multicolumn{6}{l}{\textit{\textbf{DHS+MICS}}}\\[-4pt]
\hspace{1em}M\textsubscript{\tiny d} & 0.127 & 0.747 & 0.77 & 0.541 & 0.37 \\[-4pt]
\hspace{1em}M\textsubscript{\tiny D} & 0.122 & 0.722 & 0.77 & 0.516 & 1.72 \\[-4pt]
\hspace{1em}M\textsubscript{\tiny M} & 0.123 & 0.715 & 0.77 & 0.499 & 3.56 \\[-4pt]
\hspace{1em}M\textsubscript{\tiny DM} & 0.121 & 0.712 & 0.78 & 0.516 & 6.31 \\[-4pt]
\hspace{1em}IHME\textsubscript{\tiny SPDE} & 0.129 & 0.738 & 0.79 & 0.580 & 0.50 \\[-4pt]
\hspace{1em}IHME\textsubscript{\tiny BYM2} & 0.128 & 0.741 & 0.79 & 0.571 & 0.34 \\

\addlinespace[0.3em]
\multicolumn{6}{l}{\textit{\textbf{DHS}}}\\[-4pt]
\hspace{1em}M\textsubscript{\tiny d} & 0.112 & 0.675 & 0.79 & 0.525 & 0.37 \\[-4pt]
\hspace{1em}M\textsubscript{\tiny D} & 0.111 & 0.675 & 0.79 & 0.497 & 1.72 \\[-4pt]
\hspace{1em}M\textsubscript{\tiny M} & 0.116 & 0.672 & 0.78 & 0.477 & 3.56 \\[-4pt]
\hspace{1em}M\textsubscript{\tiny DM} & 0.112 & 0.664 & 0.80 & 0.497 & 6.31 \\[-4pt]
\hspace{1em}IHME\textsubscript{\tiny SPDE} & 0.123 & 0.716 & 0.80 & 0.562 & 0.50 \\[-4pt]
\hspace{1em}IHME\textsubscript{\tiny BYM2} & 0.123 & 0.723 & 0.79 & 0.555 & 0.34 \\

\addlinespace[0.3em]
\multicolumn{6}{l}{\textit{\textbf{MICS}}}\\[-4pt]
\hspace{1em}M\textsubscript{\tiny d} & 0.145 & 0.833 & 0.74 & 0.561 & 0.37 \\[-4pt]
\hspace{1em}M\textsubscript{\tiny D} & 0.136 & 0.780 & 0.75 & 0.539 & 1.72 \\[-4pt]
\hspace{1em}M\textsubscript{\tiny M} & 0.132 & 0.767 & 0.75 & 0.526 & 3.56 \\[-4pt]
\hspace{1em}M\textsubscript{\tiny DM} & 0.133 & 0.769 & 0.76 & 0.539 & 6.31 \\[-4pt]
\hspace{1em}IHME\textsubscript{\tiny SPDE} & 0.136 & 0.766 & 0.78 & 0.602 & 0.50 \\[-4pt]
\hspace{1em}IHME\textsubscript{\tiny BYM2} & 0.134 & 0.763 & 0.78 & 0.589 & 0.34 \\[0.5em]
\bottomrule
\end{tabular}
\caption{Cluster level validation results.}
\label{tab:validationCluster1}
\end{table}

\begin{table}[ht]
\centering
\begin{tabular}{rrrrrr}
\toprule
\hspace{1em}\textbf{Approach} & \textbf{CRPS} & \textbf{80\% IS} & \textbf{80\% Cvg} & \textbf{80\% Width} & \textbf{Time (min.)} \\
\bottomrule
\addlinespace[0.3em]
\multicolumn{6}{l}{\textit{\textbf{DHS+MICS}}}\\[-4pt]
\hspace{1em}M\textsubscript{\tiny d} & 0.039 & 0.244 & 0.81 & 0.154 & 0.57 \\[-4pt]
\hspace{1em}M\textsubscript{\tiny D} & 0.038 & 0.252 & 0.78 & 0.152 & 1.74 \\[-4pt]
\hspace{1em}M\textsubscript{\tiny M} & 0.039 & 0.258 & 0.78 & 0.152 & 5.20 \\[-4pt]
\hspace{1em}M\textsubscript{\tiny DM} & 0.033 & 0.224 & 0.74 & 0.133 & 6.67 \\[-4pt]
\hspace{1em}IHME\textsubscript{\tiny SPDE} & 0.043 & 0.312 & 0.59 & 0.122 & 0.71 \\[-4pt]
\hspace{1em}IHME\textsubscript{\tiny BYM2} & 0.038 & 0.263 & 0.70 & 0.141 & 0.43 \\

\addlinespace[0.3em]
\multicolumn{6}{l}{\textit{\textbf{DHS}}}\\[-4pt]
\hspace{1em}M\textsubscript{\tiny d} & 0.044 & 0.300 & 0.81 & 0.179 & 0.57 \\[-4pt]
\hspace{1em}M\textsubscript{\tiny D} & 0.043 & 0.294 & 0.84 & 0.178 & 1.74 \\[-4pt]
\hspace{1em}M\textsubscript{\tiny M} & 0.049 & 0.333 & 0.73 & 0.179 & 5.20 \\[-4pt]
\hspace{1em}M\textsubscript{\tiny DM} & 0.042 & 0.280 & 0.79 & 0.165 & 6.67 \\[-4pt]
\hspace{1em}IHME\textsubscript{\tiny SPDE} & 0.051 & 0.381 & 0.73 & 0.154 & 0.71 \\[-4pt]
\hspace{1em}IHME\textsubscript{\tiny BYM2} & 0.048 & 0.326 & 0.70 & 0.168 & 0.43 \\

\addlinespace[0.3em]
\multicolumn{6}{l}{\textit{\textbf{MICS}}}\\[-4pt]
\hspace{1em}M\textsubscript{\tiny d} & 0.050 & 0.300 & 0.68 & 0.191 & 0.57 \\[-4pt]
\hspace{1em}M\textsubscript{\tiny D} & 0.050 & 0.306 & 0.76 & 0.189 & 1.74 \\[-4pt]
\hspace{1em}M\textsubscript{\tiny M} & 0.047 & 0.288 & 0.70 & 0.188 & 5.20 \\[-4pt]
\hspace{1em}M\textsubscript{\tiny DM} & 0.044 & 0.290 & 0.68 & 0.175 & 6.67 \\[-4pt]
\hspace{1em}IHME\textsubscript{\tiny SPDE} & 0.053 & 0.338 & 0.59 & 0.167 & 0.71 \\[-4pt]
\hspace{1em}IHME\textsubscript{\tiny BYM2} & 0.047 & 0.294 & 0.68 & 0.181 & 0.43 \\[0.5em]
\bottomrule
\end{tabular}
\caption{Admin1 level (areal) validation results.}
\label{tab:validationAreal1}
\end{table}

Cluster level validation scores and metrics are provided in Table \ref{tab:validationCluster1}, and Admin1 level (areal) validation scores and metrics in Table \ref{tab:validationAreal1}. Both sets of scores are reported for the DHS, MICS, and combined (DHS+MICS) validation data. We focus here on combined dataset CRPS as a single measure of overall predictive ability, since it is a strictly proper scoring rule, but include other metrics for interpretation purposes.

We find that, while predictions were better when using fusion models, and also when accounting for positional uncertainty in a principled way, areal predictions benefitted in particular when doing both at once. Table \ref{tab:validationAreal1} shows that, while the IHME\textsubscript{SPDE} model had the worst CRPS at 0.043 the M\textsubscript{\tiny DM} achieved the best CRPS at 0.033, the other models all had a CRPS of 0.038-0.039. The IHME approaches had the two worst areal IS, the IHME\textsubscript{SPDE} approach had the worst areal CRPS, and the IHME\textsubscript{BYM2} approach had an areal CRPS tied with the M\textsubscript{\tiny D} approach at 0.038 despite being a fusion model using both DHS and MICS datasets.

At the cluster level, Table \ref{tab:validationCluster1} shows that, while fusion modeling does tend to improve predictions relative to using only DHS data, adding MICS data matters less than accounting for positional uncertainty in a principled way. For example, the M\textsubscript{\tiny D} has better cluster level validation CRPS than M\textsubscript{\tiny d} (0.122 vs.~0.127), though still comparable, and that M\textsubscript{\tiny DM} has the best CRPS (0.121), but not much improvement compared to the M\textsubscript{\tiny D} model. The IHME models performed relatively poorly at the cluster level despite being fusion models due to their less principled approach to accounting for positional uncertainty, with a CRPS of 0.129 and 0.128 for the IHME\textsubscript{\tiny SPDE} and IHME\textsubscript{\tiny BYM2} models respectively.

While the optimal parameters are not known, the IHME approaches estimate very different covariate effects, particularly for urbanicity and population density, than the other approaches. This is consistent with the simulation study, where the IHME estimates of the parameters were significantly biased. Without accurate spatial covariate effect estimates, the IHME approaches had particular difficulty with fine scale spatial predictions.



Despite performing better than all other approaches considered at both cluster and Admin1 validation levels, the M\textsubscript{\tiny DM} approach had an average runtime of only about 7 minutes. While this is longer than the other approaches, it is certainly not prohibitive, and is still orders of magnitude faster than the runtime of \citet{wilson2021estimation}, which took approximately 52 hours to run 1000 MCMC iterations with only 398 surveyed clusters.

\FloatBarrier

\section{Conclusions} \label{sec:conclusions}

In this work, we propose a new approach capable of accounting for multiple kinds of positional error in geostatistical models, including both jittering and geomasking. It can also be performed relatively quickly compared to alternative methods suitable for such contexts such as those involving MCMC. It is more flexible in terms of the kinds of positional uncertainty allowed, the likelihood families allowed, and the method by which integration points and weights are automatically constructed based directly on the distributional form of the positional uncertainty resulting from the positional uncertainty distribution. While the focus of this paper is to use the proposed approach to build a fusion model for DHS and MICS data exhibiting jittered and geomasked point locations respectively, the approach's flexibility makes it applicable in other contexts with positional uncertainty as well. While previous limitations prevented such fusion models from being developed, the proposed model enables the use of more survey data, resulting in improved parameter estimates and predictions.

We demonstrate via simulation study and validation that using this method to account for positional anonymization can improve areal and point level predictions as well as parameter estimates, reducing attenuation bias, and improving estimate precision. We recommend the proposed method particularly in cases where the modeler wishes to model DHS and MICS surveys jointly, where continuously varying spatial covariates are used, and where local spatial estimates are desired. In these settings, we find that simultaneously accounting for positional uncertainty in a principled way while jointly modeling DHS and MICS data using the proposed fusion model improves fine scale and areal spatial predictions along with parameter estimates.

Fine scale spatial modeling of health and demographic indicators are an increasingly emphasized aspect of the UN sustainable development goals in part because they enable targeted monitoring, interventions, and resource allocation. Hence, methods for improving such fine scale predictions are particularly important. Model identifiability under positional uncertainty in cases where spatial random effects may vary significantly on fine spatial scales warrant further study, however. In addition, it is possible that accounting for uncertainty from urbanicity classification may improve predictions, although that is a complex problem beyond the scope of this work.

\section{Declaration of generative AI and AI-assisted technologies in the manuscript preparation process}
During the preparation of this work the authors used generative AI tools to assist with debugging code and formatting tables and plots originally developed by the authors. After using this tool/service, the authors reviewed and edited the content as needed and take full responsibility for the content of the published article.

\FloatBarrier

\bibliography{myBib}

@article{sorbye201439,
title = {Scaling intrinsic {G}aussian {M}arkov random field priors in spatial modelling},
journal = {Spatial Statistics},
volume = {8},
pages = {39-51},
year = {2014},
note = {Spatial Statistics Miami},
issn = {2211-6753},
doi = {https://doi.org/10.1016/j.spasta.2013.06.004},
url = {https://www.sciencedirect.com/science/article/pii/S2211675313000407},
author = {Sigrunn Holbek Sørbye and Håvard Rue}
}

@manual{Li:2024aa,
	author = {Zehang R Li and Bryan D Martin and Yuan Hsiao and Jessica Godwin and John Paige and Peter Gao and Jon Wakefield and Samuel J Clark and Geir-Arne Fuglstad and Andrea Riebler},
	note = {R package version 1.4.0},
	title = {SUMMER: Small-Area-Estimation Unit/Area Models and Methods for Estimation in R},
	url = {https://CRAN.R-project.org/package=SUMMER},
	year = {2024}}

@inproceedings{schubert2019faster,
	author = {Schubert, Erich and Rousseeuw, Peter J},
	booktitle = {Similarity Search and Applications: 12th International Conference, {SISAP} 2019, {Newark, NJ, USA, October} 2--4, 2019, Proceedings 12},
	organization = {Springer},
	pages = {171--187},
	title = {Faster k-medoids clustering: improving the {PAM, CLARA, and CLARANS} algorithms},
	year = {2019}}

@techreport{wu2021dhs,
	author = {Yunhan Wu and Zehang Richard Li and Benjamin K. Mayala and Houjie Wang and Peter Gao and John Paige and Geir-Arne Fuglstad and Caitlin Moe and Jessica Godwin and Rose E. Donohue and Trevor N. Croft and Jon Wakefield},
	entrysubtype = {techreport},
	institution = {{DHS} Spatial Analysis Reports No.~21},
	title = {Spatial Modeling for Subnational Administrative Level 2 Small-Area Estimation},
	url = {https://dhsprogram.com/publications/publication-SAR21-Spatial-Analysis-Reports.cfm},
	year = {2021}}

@article{gascoigne2023estimating,
	author = {Gascoigne, Connor and Smith, Theresa and Paige, John and Wakefield, Jon},
	journal = {{arXiv} preprint {arXiv}:2309.02093},
	title = {Estimating Subnational Under-Five Mortality Rates Using a Spatio-Temporal Age-Period-Cohort Model},
	year = {2023}}

@article{paige2022spatial,
	author = {Paige, John and Fuglstad, Geir-Arne and Riebler, Andrea and Wakefield, Jon},
	journal = {Spatial Statistics},
	pages = {100714},
	publisher = {Elsevier},
	title = {Spatial aggregation with respect to a population distribution: Impact on inference},
	volume = {52},
	year = {2022}}

@report{DHSspatial07,
	address = {USA:MD},
	author = {Clara R. Burgert and Josh Colston and Thea Roy and Blake Zachary},
	howpublished = {\url{https://dhsprogram.com/pubs/pdf/SAR7/SAR7.pdf}},
	note = {{DHS} Spatial Analysis Reports No. 7},
	title = {Geographic displacement procedure and georeferenced datarelease policy for the {D}emographic and {H}ealth {S}urveys},
	year = {2013}}

@article{marquez2021harmonizing,
	author = {Marquez, Neal and Wakefield, Jon},
	journal = {Statistical Methods in Medical Research},
	pages = {1187--1210},
	publisher = {{SAGE} Publications Sage {UK}: London, England},
	title = {Harmonizing child mortality data at disparate geographic levels},
	volume = {30},
	year = {2021}}

@techreport{MICS,
	address = {Minneapolis, MN},
	author = {A. Bolgrien and E. H. Boyle and M. Sobek and M. King},
	doi = {https://doi.org/10.18128/D082.V1.1},
	institution = {{IPUMS}},
	title = {{IPUMS MICS} Data Harmization Code. Version 1.1 [Stata Syntax]},
	year = {2024}}

@techreport{NMICS16,
	address = {Abuja, Nigeria},
	author = {{National Bureau of Statistics (NBS) and United Nations Children's Fund (UNICEF)}},
	institution = {National Bureau of Statistics and United Nations Children's Fund},
	title = {Multiple Indicator Cluster Survey 2016-17, Survey Findings Report},
	year = {2017}}

@article{khan2019multiple,
	author = {Khan, Shane and Hancioglu, Attila},
	journal = {Studies in Family Planning},
	pages = {279--286},
	publisher = {Wiley Online Library},
	title = {Multiple indicator cluster surveys: delivering robust data on children and women across the globe},
	volume = {50},
	year = {2019}}

@article{wilson2021estimation,
	author = {Wilson, Katie and Wakefield, Jon},
	journal = {Spatial and Spatio-temporal Epidemiology},
	pages = {100421},
	publisher = {Elsevier},
	title = {Estimation of health and demographic indicators with incomplete geographic information},
	volume = {37},
	year = {2021}}

@article{altay2022fast,
	author = {Altay, Umut and Paige, John and Riebler, Andrea and Fuglstad, Geir-Arne},
	journal = {arXiv preprint arXiv:2202.11035},
	title = {Fast geostatistical inference under positional uncertainty: Analysing {DHS} household survey data},
	year = {2022}}

@article{altay2024impact,
	author = {Altay, Umut and Paige, John and Riebler, Andrea and Fuglstad, Geir-Arne},
	journal = {Statistical Modelling},
	pages = {1471082X231219847},
	publisher = {{SAGE} Publications Sage India: New Delhi, India},
	title = {Impact of jittering on raster-and distance-based geostatistical analyses of {DHS} data},
	year = {2024}}

@book{perez2013guidelines,
	author = {Perez-Heydrich, Carolina and Warren, Joshua L and Burgert, Clara R and Emch, Michael},
	publisher = {{ICF} International},
	title = {Guidelines on the use of {DHS} {GPS} data},
	year = {2013}}

@article{cressie2003spatial,
	author = {Cressie, Noel and Kornak, John},
	journal = {Statistical Science},
	pages = {436--456},
	publisher = {{JSTOR}},
	title = {Spatial statistics in the presence of location error with an application to remote sensing of the environment},
	year = {2003}}

@article{lbd2020mapping,
	author = {{LBD Child Growth Failure Collaborators and others}},
	journal = {Nature medicine},
	pages = {750--759},
	publisher = {Nature Publishing Group {US} New York},
	title = {Mapping local patterns of childhood overweight and wasting in low-and middle-income countries between 2000 and 2017},
	volume = {26},
	year = {2020}}

@online{pop,
	author = {{World Pop}},
	title = {Open Spatial Demographic Data and Research},
	url = {https://hub.worldpop.org},
	urldate = {2022-09-06},
	year = 2022}

@article{weiss2018global,
	author = {Weiss, D Jꎬ and Nelson, Andy and Gibson, HS and Temperley, W and Peedell, Stephen and Lieber, A and Hancher, M and Poyart, Eduardo and Belchior, Sim{\~a}o and Fullman, Nancy and others},
	journal = {Nature},
	number = {7688},
	pages = {333--336},
	publisher = {Nature Publishing Group},
	title = {A global map of travel time to cities to assess inequalities in accessibility in 2015},
	volume = {553},
	year = {2018}}

@online{riverLake,
	author = {{Natural Earth}},
	title = {Rivers + lake centerlines},
	url = {http://www.naturalearthdata.com/downloads/10m-physical-vectors/10m-rivers-lake-centerlines},
	urldate = {2022-09-06},
	year = 2012}

@online{elev,
	author = {{National Oceanic and Atmospheric Administration}},
	title = {National Centers for Environmental Information},
	url = {https://www.ngdc.noaa.gov/mgg/topo/gltiles.html},
	urldate = {2022-09-06},
	year = 2022}

@article{godwin2021space,
	author = {Godwin, Jessica and Wakefield, Jon},
	journal = {Statistics in Medicine},
	pages = {1593--1638},
	publisher = {Wiley Online Library},
	title = {Space-time modeling of child mortality at the {A}dmin-2 level in a low and middle income countries context},
	volume = {40},
	year = {2021}}

@techreport{NigeriaDHS:18,
	address = {Abuja, Nigeria and Rockville, Maryland, {USA}},
	author = {{National Population Comission (NPC) [Nigeria] and ICF}},
	institution = {{NPC and ICF}},
	title = {Nigeria Demographic and Health Survey 2018},
	year = {2019}}

@techreport{MalawiDHS:15,
	address = {Zomba, Malawi, and Rockville, Maryland, USA},
	author = {{National Statistical Office (NSO) [Malawi] and ICF}},
	institution = {{NSO and ICF}},
	title = {Malawi Demographic and Health Survey 2015-16},
	year = {2017}}

@techreport{BurundiDHS:10,
	address = {Bujumbura, Burundi},
	author = {{Institut de Statistiques et d'\'{E}tudes \'{E}conomiques du Burundi (ISTEEBU), Minist\`{e}re de la Sant\'{e} Publique et de la Lutte contre le Sida [Burundi] (MSPLS), et ICF International}},
	institution = {ISTEEBU, MSPLS, et ICF International},
	title = {Enqu\^{e}te D\'{e}mographique et de Sant\'{e} Burundi 2010},
	year = {2012}}

@article{local2021mapping,
	author = {{Local Burden of Disease Vaccine Coverage Collaborators and others}},
	journal = {Nature},
	pages = {415--419},
	publisher = {Nature Publishing Group},
	title = {Mapping routine measles vaccination in low-and middle-income countries},
	volume = {589},
	year = {2021}}

@article{local2020mapping,
	author = {{Local Burden of Disease Child Growth Failure Collaborators and others}},
	journal = {Nature},
	pages = {231--234},
	publisher = {Nature Publishing Group},
	title = {Mapping child growth failure across low-and middle-income countries},
	volume = {577},
	year = {2020}}

@article{paige2022bayesian,
	author = {Paige, John and Fuglstad, Geir-Arne and Riebler, Andrea and Wakefield, Jon},
	journal = {Computational Statistics \& Data Analysis},
	pages = {107503},
	publisher = {Elsevier},
	title = {Bayesian multiresolution modeling of georeferenced data: {A}n extension of `{LatticeKrig}'},
	volume = {173},
	year = {2022}}

@misc{GADM,
	author = {{Global Administrative Areas}, {{\relax (GADM)}}},
	howpublished = {http://www.gadm.org},
	title = {University of {C}alifornia, {B}erkeley},
	urldate = {27/7/2021},
	year = {2018}}

@article{dong2021space,
	author = {Dong, Tracy Qi and Wakefield, Jon},
	journal = {The Annals of Applied Statistics},
	pages = {1959--1979},
	publisher = {Institute of Mathematical Statistics},
	title = {Space-time smoothing models for subnational measles routine immunization coverage estimation with complex survey data},
	volume = {15},
	year = {2021}}

@article{devarajan2013africa,
	author = {Devarajan, Shantayanan},
	journal = {Review of Income and Wealth},
	pages = {S9--S15},
	publisher = {Wiley Online Library},
	title = {Africa's statistical tragedy},
	volume = {59},
	year = {2013}}

@book{jerven2013poor,
	author = {Jerven, Morten},
	publisher = {Cornell University Press},
	title = {Poor numbers: how we are misled by {A}frican development statistics and what to do about it},
	year = {2013}}

@article{sandefur2015political,
	author = {Sandefur, Justin and Glassman, Amanda},
	journal = {The Journal of Development Studies},
	pages = {116--132},
	publisher = {Taylor \& Francis},
	title = {The political economy of bad data: {E}vidence from {A}frican survey and administrative statistics},
	volume = {51},
	year = {2015}}

@article{wagner:etal:18,
	author = {Wagner, Zachary and Heft-Neal, Sam and Bhutta, Zulfiqar A and Black, Robert E and Burke, Marshall and Bendavid, Eran},
	journal = {The Lancet},
	publisher = {Elsevier},
	title = {Armed conflict and child mortality in {A}frica: a geospatial analysis},
	year = {2018}}

@article{simpson:etal:17,
	author = {Simpson, D. and Rue, H. and Riebler, A. and Martins, T.G. and S{\o}rbye, S.H.},
	journal = {Statistical Science},
	optnumber = {1},
	pages = {1--28},
	publisher = {Institute of Mathematical Statistics},
	title = {Penalising model component complexity: A principled, practical approach to constructing priors (with discussion)},
	volume = {32},
	year = {2017}}

@manual{sdgsWeb,
	address = {\url{http://sustainabledevelopment.un.org/owg.html}},
	author = {{United Nations}},
	month = {accessed 2024-2-16},
	title = {Sustainable Development Goals},
	year = {2020}}

@book{samplingManualDHS,
	address = {Calverton, Maryland, {USA}: {ICF} International},
	author = {{ICF International}},
	institution = {MEASURE DHS},
	title = {Demographic and Health Survey Sampling and Household Listing Manual},
	year = {2012}}

@article{rue2009approximate,
	author = {Rue, H. and Martino, S. and Chopin, N.},
	journal = {Journal of the Royal Statistical Society: Series B},
	pages = {319--392},
	publisher = {Wiley Online Library},
	title = {Approximate {B}ayesian inference for latent {G}aussian models by using integrated nested {L}aplace approximations},
	volume = {71},
	year = {2009}}

@article{riebler:etal:16,
	author = {Riebler, A. and S{\o}rbye, S.H. and Simpson, D. and Rue, H.},
	journal = {Statistical Methods in Medical Research},
	optnumber = {4},
	pages = {1145--1165},
	publisher = {{SAGE} Publications Sage {UK}: London, England},
	title = {An intuitive {B}ayesian spatial model for disease mapping that accounts for scaling},
	volume = {25},
	year = {2016}}

@article{paige2022design,
	author = {John Paige and Geir-Arne Fuglstad and Andrea Riebler and Jon Wakefield},
	journal = {Journal of Survey Statistics and Methodology},
	pages = {50--80},
	title = {Design- and Model-Based Approaches to Small-Area Estimation in a Low and Middle Income Country Context: Comparisons and Recommendations},
	volume = {10},
	year = {2022}}

@article{osgood2023statistical,
  title={A statistical review of template model builder: A flexible tool for spatial modelling},
  author={Osgood-Zimmerman, Aaron and Wakefield, Jon},
  journal={International Statistical Review},
  volume={91},
  pages={318--342},
  year={2023},
  publisher={Wiley Online Library}
}

@article{osgood:etal:18,
	author = {Osgood-Zimmerman, Aaron and Millear, Anoushka I and Stubbs, Rebecca W and Shields, Chloe and Pickering, Brandon V and Earl, Lucas and Graetz, Nicholas and Kinyoki, Damaris K and Ray, Sarah E and Bhatt, Samir and Browne, A.J. and Burstein, R. and Cameron, E. and Casey, D.C. and Deshpande, A. and Fullman, N. and Gething, P.W. and Gibson, H.S. and Henry, N.J. and Herrero, M. and Krause, L.K. and Letourneau, I.D. and Levine, A.J. and Liu, P.Y. and Longbottom, J. and Mayala, B.K. and Mosser, J.F. and Noor, A.M. and Pigott, D.M. and Piwoz, E.G. and Rao, P. and Rawat, R. and Reiner, R.C. and Smith, D.L. and Weiss, D.J. and Wiens, K.E. and Mokdad, A.H. and Lim. S.S. and Murray, C.J.L. and Kassebaum, N.J. and Hay, S.I.},
	journal = {Nature},
	optnumber = {7694},
	pages = {41-47},
	publisher = {Nature Publishing Group},
	title = {Mapping child growth failure in {A}frica between 2000 and 2015},
	volume = {555},
	year = {2018}}

@Manual{fastGHQuad,
    title = {{fastGHQuad}: Fast '{Rcpp}' Implementation of {Gauss-Hermite} Quadrature},
    author = {Alexander W Blocker},
    year = {2022},
    note = {R package version 1.0.1},
    url = {https://CRAN.R-project.org/package=fastGHQuad},
  }

@article{lumley:scott:17,
	author = {T. Lumley and A. Scott},
	journal = {Statistical Science},
	title = {Fitting regression models to survey data},
	year = {2017}}

@misc{Lumley:2023,
	author = {Thomas Lumley},
	note = {R package version 4.2-1},
	title = {survey: analysis of complex survey samples},
	year = {2023}}

@article{Lindgren:etal:11,
	author = {F. Lindgren and H. Rue and J. Lindstr\"{o}m},
	journal = {Journal of the Royal Statistical Society, Series B},
	pages = {423--498},
	title = {An explicit link between {G}aussian fields and {G}aussian {M}arkov random fields: the stochastic differential equation approach (with discussion)},
	volume = {73},
	year = {2011}}

@article{li:etal:19,
	author = {Zehang R Li and Yuan Hsiao and Jessica Godwin and Bryan D Martin and Jon Wakefield and Samuel J Clark},
	journal = {P{LoS one}},
	note = {Published January 22, 2019},
	title = {Changes in the spatial distribution of the under five mortality rate: small-area analysis of 122 {DHS} surveys in 262 subregions of 35 countries in {A}frica},
	volume = {14},
	year = {2019}}

@article{kristensen:etal:15,
	author = {Kasper Kristensen and Anders Nielsen and Casper W. Berg and Hans Skaug and Bradley M. Bell},
	journal = {Journal of Statistical Software},
	optdoi = {10.18637/jss.v070.i05},
	optnumber = {5},
	pages = {1--21},
	title = {{TMB}: {A}utomatic {D}ifferentiation and {L}aplace Approximation},
	volume = {70},
	year = {2016}}

@book{KenyaCensus:2009,
	author = {{Kenya National Bureau of Statistics, Ministry of Health/Kenya, National AIDS Control Council/Kenya, Kenya Medical Research Institute, and National Council For Population And Development/Kenya}},
	publisher = {Nairobi: Kenya National Bureau of Statistics},
	shortauthor = {KDHS},
	title = {The 2009 Kenya Population and Housing Census Volume {IC}: Population Distribution by Age, Sex, and Administrative Units},
	year = {2009}}

@book{KDHS2014,
	address = {Rockville, Maryland, USA},
	author = {{Kenya National Bureau of Statistics, Ministry of Health/Kenya, National AIDS Control Council/Kenya, Kenya Medical Research Institute, and National Council For Population And Development/Kenya}},
	shortauthor = {KDHS},
	title = {Kenya Demographic and Health Survey 2014},
	url = {http://dhsprogram.com/pubs/pdf/FR308/FR308.pdf},
	year = {2015}}

@article{horvitz:thompson:52,
	author = {D.G. Horvitz and D.J. Thompson},
	journal = {Journal of the American Statistical Association},
	pages = {663--685},
	title = {A generalization of sampling without replacement from a finite universe},
	volume = {47},
	year = {1952}}

@article{graetz:etal:18,
	author = {Graetz, Nicholas and Friedman, Joseph and Osgood-Zimmerman, Aaron and Burstein, Roy and Biehl, Molly H and Shields, Chloe and Mosser, Jonathan F and Casey, Daniel C and Deshpande, Aniruddha and Earl, Lucas and Reiner, R.C. and Ray, S.E. and Fullman, N. and Levine, A.J. and Stubbs, R.W. and Mayala, B.K. and Longbottom, J. and Browne, A.J. and Bhatt, S. and Weiss, D.J. and Gething, P. and Mokdad, A. and Lim, S.S. and Murray, C.J.L. and Gakidou, E. and Hay, S.I.},
	journal = {Nature},
	optnumber = {7694},
	pages = {48-53},
	publisher = {Nature Publishing Group},
	title = {Mapping local variation in educational attainment across {A}frica},
	volume = {555},
	year = {2018}}

@article{gomez:rue:18,
	author = {G{\'o}mez-Rubio, Virgilio and Rue, H{\aa}vard},
	journal = {Statistics and Computing},
	pages = {1033--1051},
	publisher = {Springer},
	title = {Markov chain {M}onte {C}arlo with the integrated nested {L}aplace approximation},
	volume = {28},
	year = {2018}}

@article{gneiting:raftery:07,
	author = {Gneiting, Tilmann and Raftery, Adrian E},
	journal = {Journal of the American Statistical Association},
	optnumber = {477},
	pages = {359--378},
	publisher = {Taylor \& Francis},
	title = {Strictly proper scoring rules, prediction, and estimation},
	volume = {102},
	year = {2007}}

@article{geyer2005fuzzy,
	author = {Geyer, Charles J and Meeden, Glen D},
	journal = {Statistical Science},
	pages = {358--366},
	publisher = {{JSTOR}},
	title = {Fuzzy and randomized confidence intervals and p-values},
	volume = {20},
	year = {2005}}

@article{fronterre:etal:18,
	author = {Fronterr{\`e}, Claudio and Giorgi, Emanuele and Diggle, Peter J},
	journal = {Spatial Statistics},
	note = {Available online, June 23, 2018},
	title = {Geostatistical inference in the presence of geomasking: a composite-likelihood approach},
	year = {2018}}

@article{fanshawe:diggle:11,
	author = {Fanshawe, TR and Diggle, PJ},
	journal = {Environmetrics},
	optnumber = {2},
	pages = {109--122},
	publisher = {Wiley Online Library},
	title = {Spatial prediction in the presence of positional error},
	volume = {22},
	year = {2011}}

@manual{DHSPweb,
	address = {\url{http://www.dhsprogram.com}},
	author = {USAID},
	month = {accessed 2024-2-16},
	organization = {{United States Agency for International Development}},
	title = {The DHS Program - Quality information to plan, monitor, and improve population, helath, and nutrition programs},
	year = {2024}}

@book{carroll:etal:06,
	author = {Carroll, Raymond J and Ruppert, David and Crainiceanu, Ciprian M and Stefanski, Leonard A},
	publisher = {Chapman and Hall/{CRC}},
	title = {Measurement Error in Nonlinear Models: A Modern Perspective},
	year = {2006}}

@article{burstein:etal:18,
	author = {Burstein, Roy and Wang, Haidong and Reiner Jr, Robert C and Hay, Simon I},
	journal = {{PLoS} Medicine},
	optnumber = {10},
	pages = {e1002687},
	publisher = {Public Library of Science},
	title = {Development and validation of a new method for indirect estimation of neonatal, infant, and child mortality trends using summary birth histories},
	volume = {15},
	year = {2018}}

@article{binder:83,
	author = {D.A. Binder},
	journal = {International Statistical Review},
	pages = {279--292},
	title = {On the variances of asymptotically normal estimators from complex surveys},
	volume = {51},
	year = {1983}}

@article{bakar2026bayesian,
  title={A {B}ayesian model adjusted for location perturbations in sample surveys: with application to the analysis of demographic health survey data},
  author={Bakar, Khandoker Shuvo and Baffour, Bernard and Khan, Jahidur Rahman and Ayebeng, Castro},
  journal={Journal of the Royal Statistical Society Series A: Statistics in Society},
  pages={qnag068},
  year={2026},
  publisher={Oxford University Press UK}
}

\pagebreak 
\appendix

\end{document}




\begin{frontmatter}

  \title{Supplementary materials for \\`Spatial fusion of DHS and MICS surveys\\with jittered and geomasked locations'}

\author[NRaddress]{John Paige\corref{mycorrespondingauthor}}
\cortext[mycorrespondingauthor]{Corresponding author}
\ead{jpaige@nr.no}

\author[NTNUaddress]{Geir-Arne Fuglstad}
\author[NTNUaddress]{Andrea Riebler}

\address[NRaddress]{Norsk Regnesentral, Oslo, Norway}
\address[NTNUaddress]{Department of Mathematical Sciences, NTNU, Trondheim, Norway}

\end{frontmatter}

\appendix
\renewcommand{\thesection}{S.\arabic{section}}

\section{MICS positional error integration scheme}
\label{sec:supplementMICSintegration}

In order to evaluate \eqref{eq:posErrMod}, we use a different numerical scheme for MICS data than for DHS data due to the differing positional uncertainty distributions. We will make the assumption that $\pi \big(s_c, t_c = t \big)$ and $\pi \big(y_c \mid r(t) \big)$ are both continuous in $t$ almost everywhere for each $c$. Note that, in the case of geomasking, $\pi \big(s_c, t_c = t \big) = \pi \big(s_c \vert t_c = t \big) \cdot \pi \big(t_c = t \big)$ where $\pi \big(s_c \vert t_c = t \big) = \mathbb{I}\{s_c \ni t\}$, and we assume $\pi(t)$ is proportional to population density, so the assumptions on continuity hold.

Our integration scheme can be derived by rewriting the integral in \eqref{eq:posErrMod} in terms of a sum of integrals over `integration zones' that partition the support of $\pi \big(s_c, t_c \big)$, say, $Z \subset \mathbb{R}^2$, for any fixed cluster $c$ (although $Z$ can be in more general spaces than $\mathbb{R}^2$). In our case we can think of $Z$ as the geomasked area containing $s_c$, or, more generally, as the set of possible points where the true location of cluster $c$ might be located. Rewriting the integral in \eqref{eq:posErrMod}, we get: 
\begin{align}
 \pi \big(y_c \mid r(\cdot) \big) &= \int_{Z} \pi \big(y_c \mid r(t_c) \big) \,  \pi \big(s_c | t_c \big) \pi(t_c) \ \mathrm{d}t_c \nonumber \\
 &= \sum_{k=1}^{K} \int_{Z^k} \pi \big(y_c \mid r(t_c) \big) \, \pi \big(s_c | t_c \big) \pi(t_c) \ \mathrm{d}t_c, \label{eq:intZones}
\end{align}
where $Z^1 \cup \ldots \cup Z^K = Z$ are the aforementioned integration zones forming a partition of $Z$. Note that any single $Z^k$ need not be spatially contiguous. Our goal will be to select a single integration point in each zone $Z^k$ that is `representative' in some sense of $Z^k$, and to choose each integration zone $Z^k$ carefully so that $\pi \big(y_c \mid r(t_c) \big)$ remains relatively constant throughout each zone as a function of $t_c$.

To set the integration zones and their representative points, we first construct a fine spatial grid, $G = \{g_{1}, \ldots, g_{|G|}\}$ within $Z$. We then consider the transformation of the spatial grid points to an alternative space where distances better represent variations in the integrand. For now, we assume the grid points are transformed via $\boldsymbol{h}(\cdot) = \boldsymbol{d}(\cdot)$, which is a transformation from spatial location to the vector of covariates at the location. The transformed grid points are given by $X = \{\boldsymbol{x}_1, \ldots, \boldsymbol{x}_{|G|}\} = \{\boldsymbol{h}(g_1), \ldots, \boldsymbol{h}(g_{|G|})\}$, where $|G|$ denotes the cardinality of $G$, i.e.~the number of fine grid points. We discuss other choices of $\boldsymbol{h}(\cdot)$ accounting for correlation due to spatial random effects later in this section.

We then wish to choose our set of $K$ representative points from among the $|G|$ grid points, along with their associated $K$ integration zones, by minimizing the criterion,
\begin{equation}
\boldsymbol{x}_1^*, \ldots, \boldsymbol{x}_{K}^*, X^1, \ldots, X^{K} = \underset{\boldsymbol{x}_1^*, \ldots, \boldsymbol{x}_{K}^*, X^1, \ldots, X^{K}}{\mbox{argmin}} \sum_{k=1}^{K} \sum_{\boldsymbol{h}(g) \in X^k} \pi \big(s_c | t_c = g \big) \pi(t_c = g) || \boldsymbol{x}_k^* - \boldsymbol{h}(g) || ^2,
\label{eq:kMedCriterionGeneral}
\end{equation}
under the constraints that $X^1 \cup \ldots \cup X^{K} = X$ form a partition of the transformed fine grid, $X$, and that $\boldsymbol{x}_k^* \in X^k$ for $k=1,\ldots, K$.

Now we can back-transform to the spatial domain, taking $\tilde{t}_{ck} \equiv \boldsymbol{h}^{-1}(\boldsymbol{x}_1^*), \ldots, t_{cK} \equiv \boldsymbol{h}^{-1}(\boldsymbol{x}_K^*)$, and defining a partition of the fine spatial grid $G^1 \cup \ldots \cup G^{K} = G$, where each $G^k$ is the preimage of $X^k$ under $\boldsymbol{h}(\cdot)$, so that $G^k \equiv \{g : \boldsymbol{h}(g) \in X^k\}$ for $k=1,\ldots K$.  By the definition of the optimization criterion in \eqref{eq:kMedCriterionGeneral}, each $\tilde{t}_{ck}$ is chosen to have similar covariate values to the other fine grid points in $G^k$, which means that the covariate values of each of the points in $G^k$ tend to be similar, with $\tilde{t}_{ck}$ being the most representative point in $G^k$.

Note that if $\boldsymbol{h}$ is not invertible so that $\boldsymbol{h}^{-1}(\boldsymbol{x}_k^*) = \{g_{i_1}, \ldots, g_{i_k}\}$ for some $k$, then we can set $\tilde{t}_{ck}$ to be any element of $\boldsymbol{h}^{-1}(\boldsymbol{x}_k^*)$ arbitrarily.

In the case of geomasked data, \eqref{eq:kMedCriterionGeneral} then simplifies to: 
\begin{equation}
\boldsymbol{x}_1^*, \ldots, \boldsymbol{x}_{K}^*, X^1, \ldots, X^{K} = \underset{\boldsymbol{x}_1^*, \ldots, \boldsymbol{x}_{K}^*, X^1, \ldots, X^{K}}{\mbox{argmin}} \sum_{k=1}^{K} \sum_{\boldsymbol{h}(g) \in X^k} q(g) || \boldsymbol{x}_k^* - \boldsymbol{h}(g) || ^2.
\label{eq:kMedCriterion}
\end{equation}
Here, $q(g)$ is the population density associated with spatial location $g$, and the criterion is the same for all clusters in the same geomasked area. Hence, while \eqref{eq:kMedCriterionGeneral} must be minimized for each cluster, \eqref{eq:kMedCriterion} only needs to be minimized for each geomasked area, leading to substantial computational savings. The $K$-medoids algorithm and the related partitioning around medoids (PAM) algorithm \citep{schubert2019faster} were built to solve minimization problems taking the form of \eqref{eq:kMedCriterionGeneral} and \eqref{eq:kMedCriterion}.

We then define the integration weights $\alpha_{ck}$, $k=1,\ldots,K$ to be proportional to the average joint density of $s_c$ and $t_c$ associated with the $k$-th representative point (medoid) divided by the joint positional density evaluated at the medoid, i.e., 
\begin{equation}
\alpha_{ck} = \frac{1}{\pi \big(s_c | t_c = \tilde{t}_{ck} \big) \pi(t_c = \tilde{t}_{ck})} \sum_{g \in G^k} \pi \big(s_c | t_c = g \big) \pi(t_c = g).
\label{eq:intWeightsGeneral}
\end{equation}
In the case of the geomasked MICS data, this simplifies to the average population density associated with medoid $k$ divided by the population density evaluated at the medoid: 
\begin{equation}
\alpha_{ck} = \frac{1}{q(\tilde{t}_{ck})} \sum_{g \in G^k} q(g).
\label{eq:intWeights}
\end{equation}
Just as for the integration points, in the case of geomasking the integration weights need only be calculated once for each of the geomasked areas rather than once for each cluster.

If $\pi \big(y_c \mid r(t_c) \big)$ were constant as a function of $t_c$ in each zone $Z^k$, then for any $k=1,\ldots,K$, we can plug in the expression for $\alpha_{ck}$ from \eqref{eq:intWeightsGeneral} back into \eqref{eq:intZones} to get:
\begin{align*}
\int_{Z^k} \pi \big(y_c \mid r(t_c) \big) \, \pi \big(s_c | t_c \big) \pi(t_c) \ \mathrm{d}t_c &= \pi \big(y_c \mid r(\tilde{t}_{ck}) \big) \int_{Z^k} \, \pi \big(s_c | t_c \big) \pi(t_c) \ \mathrm{d}t_c \\
&\approx \pi \big(y_c \mid r(\tilde{t}_{ck}) \big) \sum_{g \in G^k} \pi \big(s_c | t_c = g \big) \pi(t_c = g) \\
&= \alpha_{ck} \, \pi \big(y_c | r(\tilde{t}_{ck}) \big) \, \pi \big(s_c | \tilde{t}_{ck} \big) \, \pi \big(\tilde{t}_{ck} \big), 
\end{align*}
and so
$$ \sum_{k = 1}^{K} \int_{Z^k} \pi \big(y_c \mid r(t_c) \big) \, \pi \big(s_c | t_c \big) \pi(t_c) \ \mathrm{d}t_c \approx \sum_{k = 1}^{K} \alpha_{ck} \, \pi \big(y_c | r(\tilde{t}_{ck}) \big) \, \pi \big(s_c | \tilde{t}_{ck} \big) \, \pi \big(\tilde{t}_{ck} \big), $$
yielding the integration scheme in \eqref{eq:intSchemeDHS}. Our approximation is therefore accurate when $\pi \big(y_c \mid r(t_c) \big)$ varies little for $t_c \in Z^k$, and when $\int_{Z^k} \, \pi \big(s_c | t_c \big) \pi(t_c) \ \mathrm{d}t \approx \sum_{g \in G^k} \pi \big(s_c | t_c = g \big) \pi(t_c=g)$. The former will be the case for small enough $Z^k$ provided $\pi \big(y_c \mid r(t_c) \big)$ is continuous in $t_c$ almost everywhere in $Z^k$. Importantly, $\pi \big(y_c \mid r(t_c) \big)$ could be discontinuous at the boundaries of $Z^k$, as can be the case for the MICS data since $Z$ is chosen the be the urban or rural part of an area, and urbanicity, which influences $\pi \big(y_c \mid r(t_c) \big)$, is discontinuous. Even if $Z$ itself is not chosen in a way to account for such discontinuities in $r(t_c)$, for large enough $K$, the clustering algorithm will choose the $Z^k$ automatically so that the discontinuities do not occur within them.

Note that spatial information can be incorporated into this integration scheme as well. For a continuous spatial model, for example, the same integration scheme can be applied using transformation $\boldsymbol{d}^*(s; \, \lambda) = (\lambda s^T \ \boldsymbol{d}(s)^T)^T$ in lieu of $\boldsymbol{d}(s)$, since this encourages integration points to be more spread out spatially within $Z$, and $\lambda$ can be chosen, for example, as the prior effective range divided by the diameter of the spatial domain, $\mathcal{D}$, or to match the range of the covariates. The parameter $\lambda$ can also be increased until the chosen sets $G^1, \ldots, G^{K^{\tiny \mbox{MICS}}}$ are fully connected if necessary. For an areal spatial model with spatial random effects occurring at finer scales than the positional uncertainty distribution (i.e.~finer than the geomasked area), one could set $\boldsymbol{d}^*(s; \, \lambda) = (\lambda \boldsymbol{a}(A^*[s])^T \ \boldsymbol{d}(s)^T)^T$, where $A^*[s]$ is the index of the specific areal spatial effect containing $s$, and $\boldsymbol{a}(\cdot)$ could be a function returning the spatial coordinates of the centroid of the area. The parameter $\lambda$ could be chosen in ways similar to the continuous spatial case, or in order to ensure that there are sufficient integration points in each modeled area.

\subsection{Constructing the integration scheme: Illustration for FCT Abuja}
\begin{figure}
\center
\vspace{-.2in}
\includegraphics[width=0.47\textwidth]{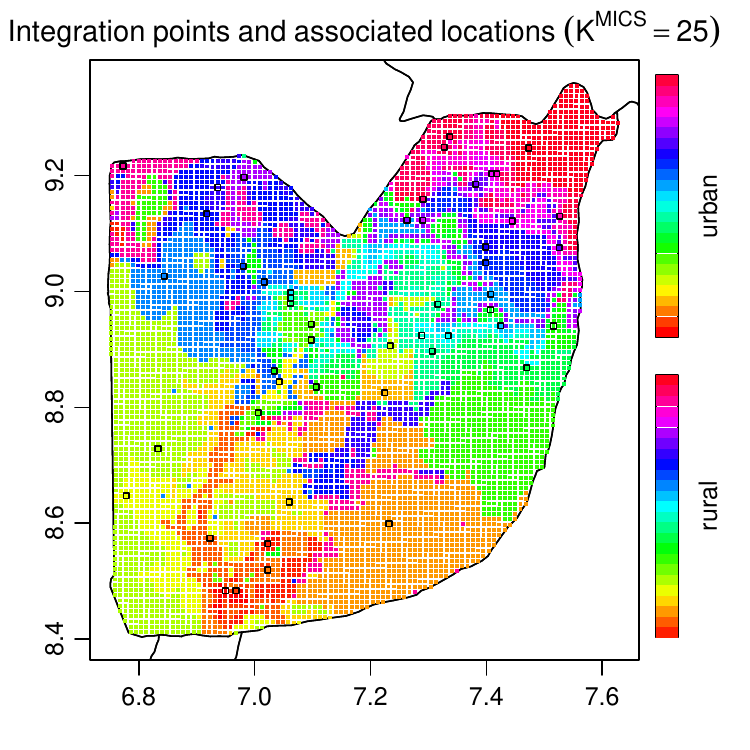} \hspace{-.22in} \includegraphics[width=0.47\textwidth]{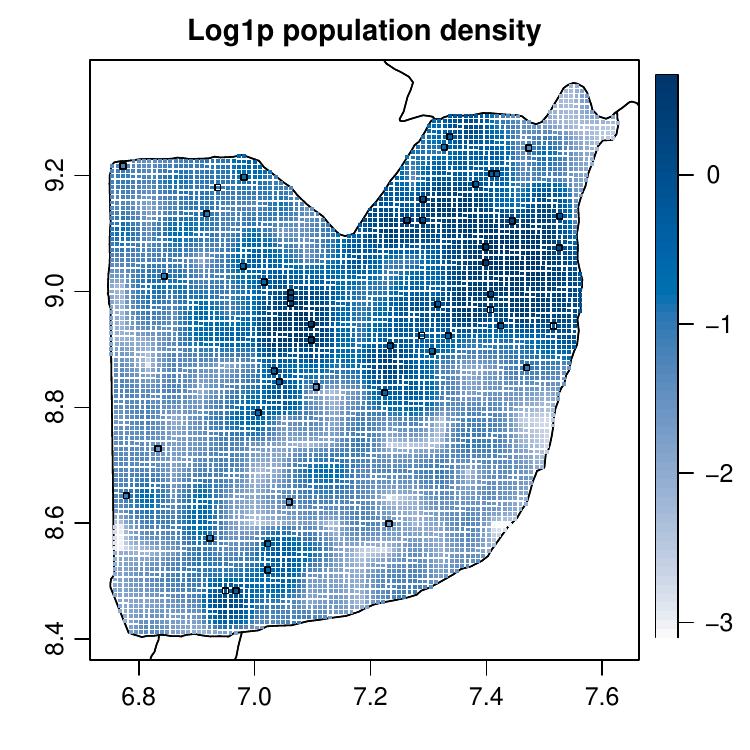} \\[-.15in]
\includegraphics[width=0.45\textwidth]{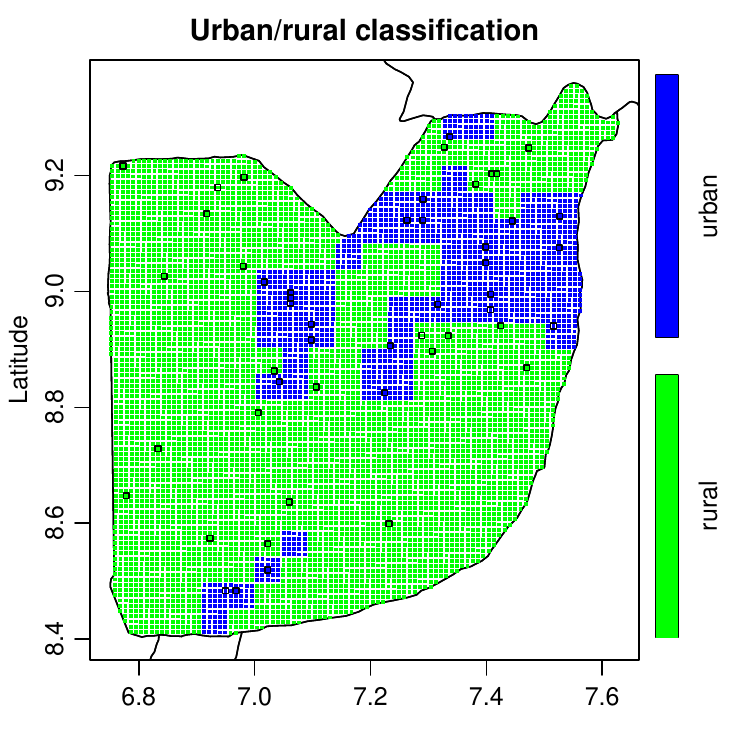} \includegraphics[width=0.45\textwidth]{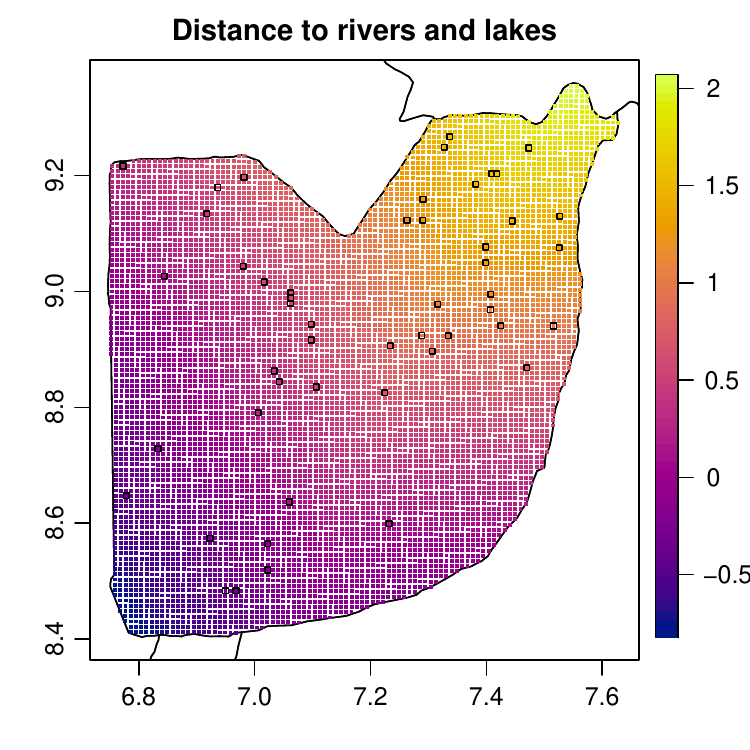} \\[-.15in]
\includegraphics[width=0.45\textwidth]{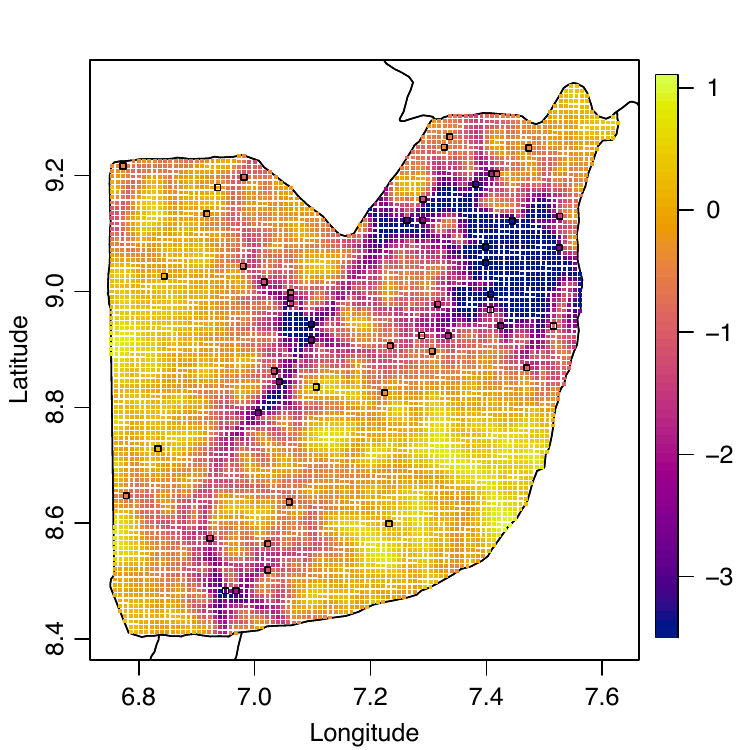} \includegraphics[width=0.45\textwidth]{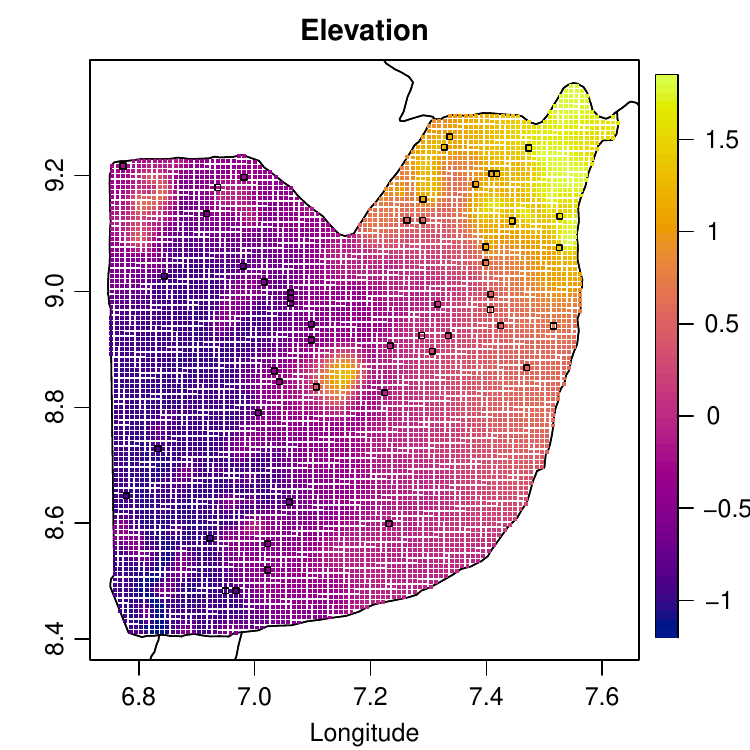}
\vspace{-.1in}
\caption{MICS integration points plotted as black squares surrounding associated fine grid points with nonzero population. Fine grid points are colored by associated integration points (top left), normalized $\log(1+\mbox{population})$ (top right), urban/rural classifications (middle left), normalized distance to rivers and lakes (middle right), normalized $\log(1+\mbox{healthcare inaccessibility})$ (bottom left), and normalized $\sqrt{\mbox{elevation}}$ (bottom right) for $K = 25$ integration points in urban and rural parts of FCT Abuja.}
\label{fig:intIllustration}
\end{figure}

\begin{figure}
\center
\includegraphics[width=0.5\textwidth]{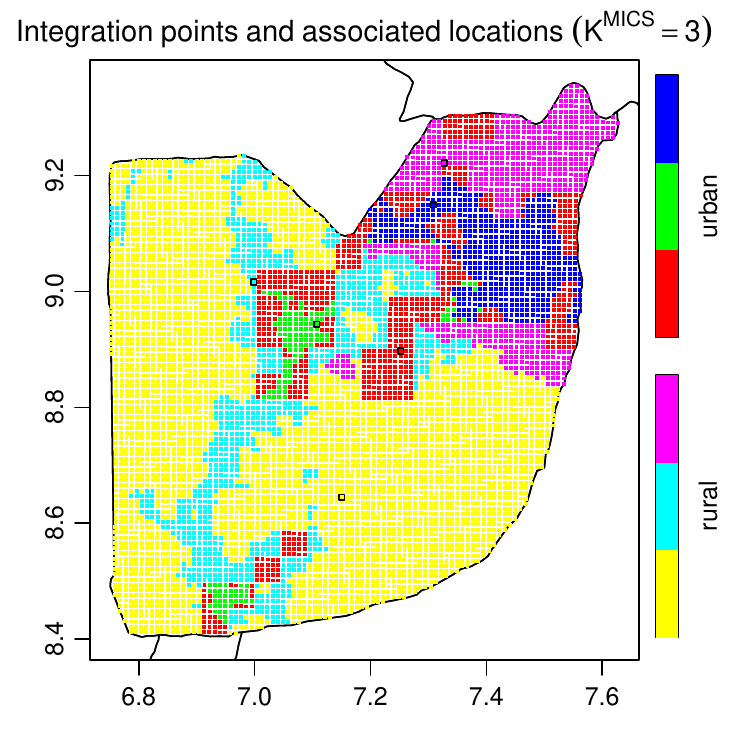} \hspace{-.15in} \includegraphics[width=0.5\textwidth]{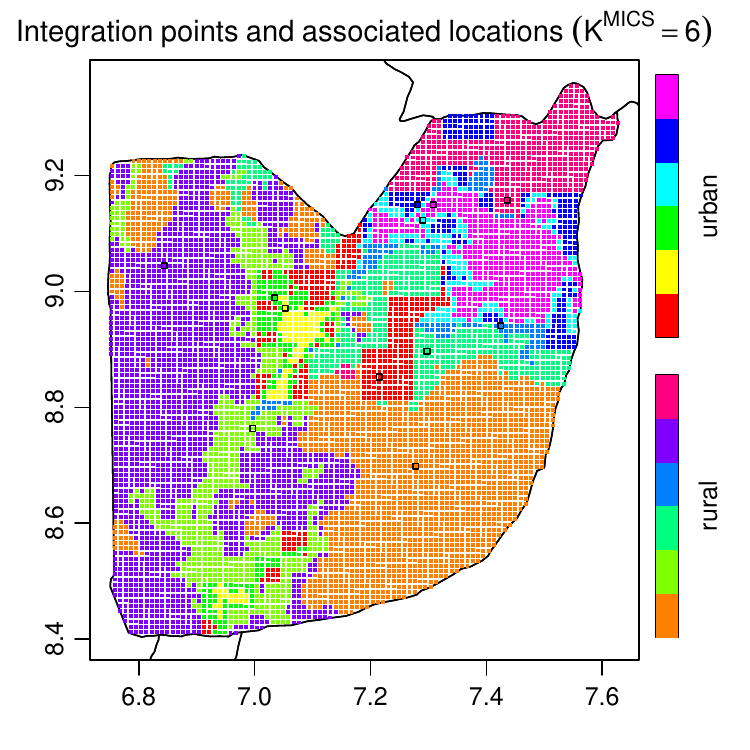} \\[-.05in]
\includegraphics[width=0.5\textwidth]{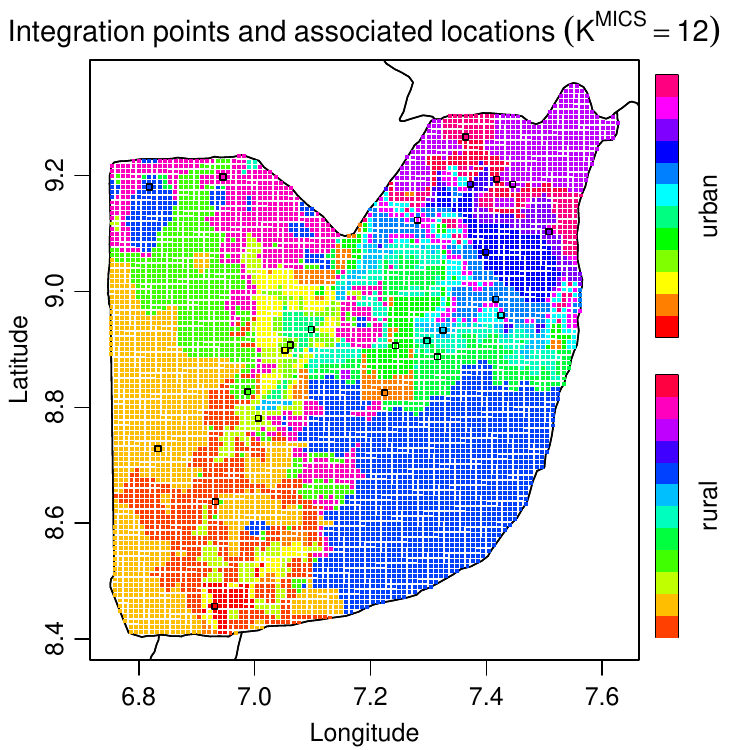} \hspace{-.15in} \includegraphics[width=0.5\textwidth]{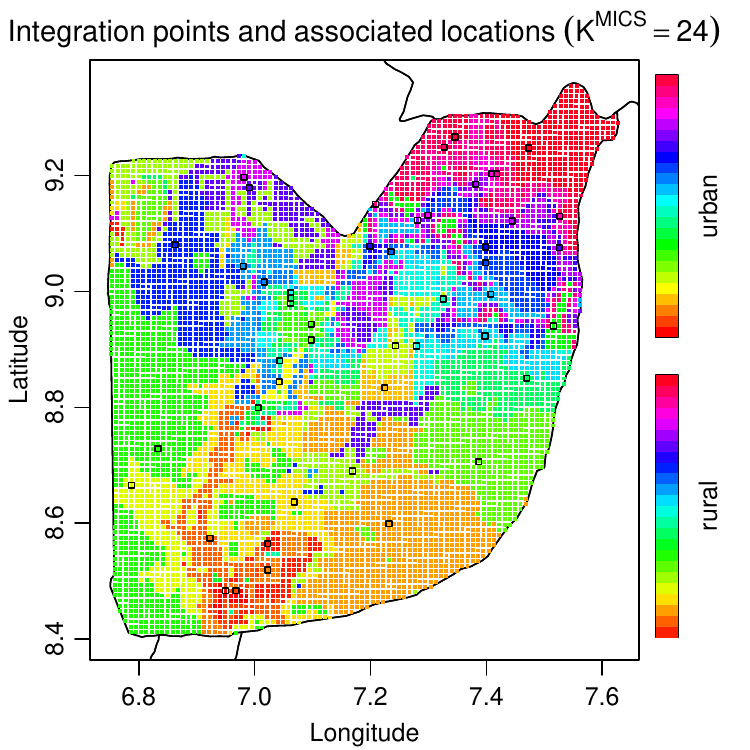}
\vspace{-.1in}
\caption{Integration points plotted as black squares surrounding associated fine grid points colored by associated integration point for $K = 3$, 6, 12, and 24 for FCT Abuja.}
\label{fig:intIllustrationInK}
\end{figure}

To illustrate selecting the MICS integration points for FCT Abuja, we first choose $K = 25$, and select a fine grid with 1km resolution. The resulting fine grid and integration points are depicted along with covariates in Figure \ref{fig:intIllustration}. In the Figure, the integration points are chosen to be representative of the different combinations of covariate values throughout the area. As $K$ increases, the grouping of fine grid integration points becomes more refined as Illustrated in Figure \ref{fig:intIllustrationInK} for $K = 3$, 6, 12, and 24. Computation and memory requirements increase linearly with the number of integration points per sampled cluster, but the accuracy of the integration improves as well. We also compare the integration scheme for $K=25$ with raster based covariates used in the analysis in Figure \ref{fig:intIllustration}.

\section{PC prior for proportion of structured variance}

The PC prior for the BYM2 parameter representing the proportion of structured variance takes the form:
\begin{align*}
f(\phi) &= \lambda \exp(-\lambda d(\phi)) \left| \frac{\partial d(\phi)}{\partial \phi} \right|\\
d(\phi) &= \sqrt{2 \mbox{KLD}(\phi)},
\end{align*}
where $d(\cdot)$ is the Kullback Liebler Divergence (KLD) distance, and the KLD is:
\begin{align*}
\mbox{KLD}(\phi) &= \frac{1}{2} \left[ \mbox{tr}\left((1-\phi) I + \phi \boldsymbol{Q}_{\star}^{-1} \right) - n - \log \left|(1-\phi)I + \phi \boldsymbol{Q}_{\star}^{-1} \right| \right] \\
&= \frac{1}{2} \left[ n(1-\phi) + \phi \mbox{tr}(\boldsymbol{Q}_{\star}^{-1}) - n - \log \left|(1-\phi)I + \phi \boldsymbol{Q}_{\star}^{-1} \right| \right] \\
&= \frac{n \phi}{2} \left( \frac{1}{n} \mbox{tr}(\boldsymbol{Q}_{\star}^{-1}) - 1 \right) - \frac{1}{2} \log \left|(1-\phi)I + \phi \boldsymbol{Q}_{\star}^{-1} \right| \\
&= \frac{n \phi}{2} \left( \left( \frac{1}{n} \sum_{i=1}^n \tilde{\gamma}_i \right) - 1 \right) - \frac{1}{2} \log \prod_{i=1}^n(1 + (\tilde{\gamma}_i - 1) \phi) \\
&= \frac{n \phi}{2} \left( \left( \frac{1}{n} \sum_{i=1}^n \tilde{\gamma}_i \right) - 1 \right) - \frac{1}{2} \sum_{i=1}^n \log(1 + (\tilde{\gamma}_i - 1) \phi).
\end{align*}
Here, $\tilde{\gamma}_i$ is an eigenvalue for $\boldsymbol{Q}_{\star}^{-1}$:
\begin{align*}
\tilde{\gamma}_i &= 1/\gamma_i \quad \text{for $\gamma_i$ eigenvalues of $\boldsymbol{Q}_{\star}$}.
\end{align*}
The PC prior Jacobian factor depends on the derivative of the KLD with respect to $\phi$, which is:
\begin{align*}
\left| \frac{\partial \mbox{KLD}(\phi)}{\partial \phi} \right| &= \left| \frac{\partial }{\partial \phi} \left[ \frac{n \phi}{2} \left( \left( \frac{1}{n} \sum_{i=1}^n \tilde{\gamma}_i \right) - 1 \right) - \frac{1}{2} \sum_{i=1}^n \log(1 + (\tilde{\gamma}_i - 1) \phi) \right] \right| \\
&= \left| \frac{n}{2} \left( \left( \frac{1}{n} \sum_{i=1}^n \tilde{\gamma}_i \right) - 1 \right) - \frac{1}{2} \sum_{i=1}^n  \frac{\partial }{\partial \phi} \log(1 + (\tilde{\gamma}_i - 1) \phi) \right| \\
&= \left| \frac{1}{2} \sum_{i=1}^n \tilde{\gamma}_i - \frac{n}{2} - \frac{1}{2} \sum_{i=1}^n  \frac{\tilde{\gamma}_i - 1}{1 + (\tilde{\gamma}_i - 1) \phi} \right|.
\end{align*}
We therefore find the Jacobian factor is:
\begin{align*}
\left| \frac{\partial d(\phi)}{\partial \phi} \right| &= \frac{1}{2} \frac{1}{\sqrt{2 \mbox{KLD}(\phi)}} \left|  \frac{\partial 2\mbox{KLD}(\phi)}{\partial \phi} \right| \\
&= \frac{1}{2\sqrt{2 \mbox{KLD}(\phi)}} \left| \sum_{i=1}^n \tilde{\gamma}_i - n - \sum_{i=1}^n  \frac{\tilde{\gamma}_i - 1}{1 + (\tilde{\gamma}_i - 1) \phi} \right|.
\end{align*}
The PC prior for $\phi$ is therefore,
\begin{equation}
f(\phi) = \lambda \exp(-\lambda d(\phi)) \frac{1}{2\sqrt{2 \mbox{KLD}(\phi)}} \left| \sum_{i=1}^n \tilde{\gamma}_i - n - \sum_{i=1}^n  \frac{\tilde{\gamma}_i - 1}{1 + (\tilde{\gamma}_i - 1) \phi} \right|,
\end{equation}
with cdf, 
\begin{equation}
F(\phi) = 1 - \exp(-\lambda d(\phi)).
\end{equation}

\FloatBarrier

\section{Additional simulation results} \label{sec:addSim}

\FloatBarrier

\begin{table}[ht]
\centering
\resizebox{1\linewidth}{!}{
\begin{tabular}{rrrrrrr}
  \hline
Model & Bias & MSE & CRPS & 95\% IS & 95\% Cvg & 95\% CI Width \\
  \hline
M\textsubscript{\tiny d} & 0.004 & 0.002 & 0.022 & 0.182 & 0.93 & 0.146 \\[-4pt]
M\textsubscript{\tiny D} & -0.002 & 0.001 & 0.022 & 0.188 & 0.93 & 0.144 \\[-4pt]
M\textsubscript{\tiny M} & -0.001 & 0.001 & 0.020 & 0.168 & 0.94 & 0.132 \\[-4pt]
M\textsubscript{\tiny DM} & -0.002 & 0.001 & 0.015 & 0.131 & 0.93 & 0.100 \\[-4pt]
IHME\textsubscript{\tiny SPDE} & 0.004 & 0.003 & 0.036 & 0.608 & 0.61 & 0.104 \\[-4pt]
IHME\textsubscript{\tiny BYM2} & 0.002 & 0.001 & 0.016 & 0.136 & 0.97 & 0.124 \\
   \hline
\end{tabular}
}
\caption{BYM2 simulation study Admin1 level prediction scores and metrics.}
\label{tab:sim1BYM2}
\end{table}

\begin{table}[ht!]
\centering
\renewcommand{\arraystretch}{0.85}
\resizebox{1.1\linewidth}{!}{
\begin{tabular}{rrrrrrr}
\toprule
 & \textbf{Bias} & \textbf{MSE} & \textbf{CRPS} & \textbf{95\% IS} & \textbf{95\% Cvg} & \textbf{95\% Width} \\
\toprule
    \multicolumn{7}{l}{\textbf{M\textsubscript{\tiny d}}} \\[-4pt]
  (Int) & 0.216 & 0.083 & 0.202 & 4.295 & 0.34 & 0.355 \\[-4pt]
  Urban & -0.026 & 0.015 & 0.071 & 0.556 & 0.94 & 0.461 \\[-4pt]
  Healthcare inaccessibility & -0.022 & 0.003 & 0.032 & 0.282 & 0.90 & 0.195 \\[-4pt]
  Elevation & -0.011 & 0.007 & 0.049 & 0.355 & 0.90 & 0.237 \\[-4pt]
  Dist.~rivers \& lakes & -0.004 & 0.007 & 0.049 & 0.488 & 0.78 & 0.218 \\[-4pt]
  Population & -0.152 & 0.037 & 0.124 & 1.581 & 0.58 & 0.387 \\
    \multicolumn{7}{l}{\textbf{M\textsubscript{\tiny D}}} \\[-4pt]
  (Int) & 0.017 & 0.045 & 0.129 & 1.921 & 0.68 & 0.410 \\[-4pt]
  Urban & 0.043 & 0.020 & 0.082 & 0.706 & 0.88 & 0.461 \\[-4pt]
  Healthcare inaccessibility & 0.003 & 0.006 & 0.043 & 0.427 & 0.87 & 0.235 \\[-4pt]
  Elevation & -0.011 & 0.009 & 0.060 & 0.535 & 0.78 & 0.241 \\[-4pt]
  Dist.~rivers \& lakes & -0.005 & 0.008 & 0.054 & 0.613 & 0.73 & 0.224 \\[-4pt]
  Population & -0.021 & 0.024 & 0.089 & 0.813 & 0.87 & 0.475 \\
    \multicolumn{7}{l}{\textbf{M\textsubscript{\tiny M}}} \\[-4pt]
  (Int) & -0.017 & 0.074 & 0.160 & 1.528 & 0.82 & 0.816 \\[-4pt]
  Urban & 0.040 & 0.084 & 0.169 & 1.402 & 0.89 & 0.861 \\[-4pt]
  Healthcare inaccessibility & 0.010 & 0.067 & 0.158 & 1.835 & 0.79 & 0.748 \\[-4pt]
  Elevation & -0.012 & 0.016 & 0.073 & 0.715 & 0.90 & 0.467 \\[-4pt]
  Dist.~rivers \& lakes & -0.003 & 0.015 & 0.068 & 0.642 & 0.91 & 0.379 \\[-4pt]
  Population & 0.007 & 0.147 & 0.225 & 2.320 & 0.85 & 1.246 \\
    \multicolumn{7}{l}{\textbf{M\textsubscript{\tiny DM}}} \\[-4pt]
  (Int) & -0.003 & 0.040 & 0.124 & 1.941 & 0.64 & 0.359 \\[-4pt]
  Urban & 0.039 & 0.011 & 0.061 & 0.459 & 0.92 & 0.375 \\[-4pt]
  Healthcare inaccessibility & 0.007 & 0.006 & 0.044 & 0.530 & 0.86 & 0.234 \\[-4pt]
  Elevation & -0.015 & 0.010 & 0.062 & 0.649 & 0.71 & 0.235 \\[-4pt]
  Dist.~rivers \& lakes & -0.007 & 0.008 & 0.054 & 0.614 & 0.73 & 0.214 \\[-4pt]
  Population & -0.006 & 0.022 & 0.085 & 0.871 & 0.84 & 0.458 \\
    \multicolumn{7}{l}{\textbf{IHME\textsubscript{\tiny SPDE}}} \\[-4pt]
  (Int) & 0.337 & 0.142 & 0.243 & 2.173 & 0.70 & 0.330 \\[-4pt]
  Urban & 0.448 & 0.204 & 0.414 & 13.522 & 0.00 & 0.233 \\[-4pt]
  Healthcare inaccessibility & 0.004 & 0.000 & 0.009 & 0.091 & 0.99 & 0.089 \\[-4pt]
  Elevation & -0.005 & 0.001 & 0.015 & 0.145 & 0.96 & 0.135 \\[-4pt]
  Dist.~rivers \& lakes & -0.009 & 0.001 & 0.021 & 0.177 & 0.96 & 0.130 \\[-4pt]
  Population & -0.432 & 0.187 & 0.408 & 14.114 & 0.00 & 0.166 \\
    \multicolumn{7}{l}{\textbf{IHME\textsubscript{\tiny BYM2}}} \\[-4pt]
  (Int) & 0.411 & 0.206 & 0.377 & 11.896 & 0.08 & 0.252 \\[-4pt]
  Urban & 0.440 & 0.197 & 0.406 & 13.084 & 0.00 & 0.238 \\[-4pt]
  Healthcare inaccessibility & 0.004 & 0.000 & 0.008 & 0.088 & 1.00 & 0.088 \\[-4pt]
  Elevation & -0.006 & 0.000 & 0.012 & 0.106 & 1.00 & 0.106 \\[-4pt]
  Dist.~rivers \& lakes & -0.001 & 0.000 & 0.012 & 0.111 & 0.99 & 0.106 \\[-4pt]
  Population & -0.447 & 0.201 & 0.424 & 14.798 & 0.00 & 0.162 \\
  \bottomrule
\end{tabular}
}
\vspace{-6pt}
\caption{SPDE simulation study parameter estimate scores and metrics with respect to true data generating process parameters. Parameters not necessarily interpretable in the same way under different models, so `best' parameters for prediction are technically unknown for scoring purposes.}
\label{tab:sim1par}
\end{table}

\begin{table}[ht!]
\centering
\renewcommand{\arraystretch}{0.85}
\resizebox{1.1\linewidth}{!}{
\begin{tabular}{rrrrrrr}
\toprule
 & \textbf{Bias} & \textbf{MSE} & \textbf{CRPS} & \textbf{95\% IS} & \textbf{95\% Cvg} & \textbf{95\% Width} \\
\toprule
    \multicolumn{7}{l}{\textbf{M\textsubscript{\tiny d}}} \\[-4pt]
  (Int) & 0.203 & 0.048 & 0.157 & 2.272 & 0.41 & 0.348 \\[-4pt]
  Urban & -0.004 & 0.012 & 0.062 & 0.545 & 0.94 & 0.448 \\[-4pt]
  Healthcare inaccessibility & -0.018 & 0.002 & 0.027 & 0.233 & 0.95 & 0.189 \\[-4pt]
  Elevation & 0.009 & 0.004 & 0.034 & 0.299 & 0.95 & 0.238 \\[-4pt]
  Dist.~rivers \& lakes & -0.002 & 0.003 & 0.030 & 0.238 & 0.96 & 0.224 \\[-4pt]
  Population & -0.154 & 0.032 & 0.116 & 1.358 & 0.67 & 0.377 \\[-4pt]
    \multicolumn{7}{l}{\textbf{M\textsubscript{\tiny D}}} \\[-4pt]
  (Int) & 0.004 & 0.011 & 0.060 & 0.491 & 0.94 & 0.414 \\[-4pt]
  Urban & 0.065 & 0.020 & 0.079 & 0.752 & 0.92 & 0.458 \\[-4pt]
  Healthcare inaccessibility & 0.004 & 0.004 & 0.034 & 0.284 & 0.97 & 0.232 \\[-4pt]
  Elevation & 0.012 & 0.005 & 0.038 & 0.348 & 0.91 & 0.251 \\[-4pt]
  Dist.~rivers \& lakes & -0.001 & 0.003 & 0.033 & 0.272 & 0.94 & 0.231 \\[-4pt]
  Population & -0.027 & 0.016 & 0.071 & 0.587 & 0.91 & 0.466 \\[-4pt]
    \multicolumn{7}{l}{\textbf{M\textsubscript{\tiny M}}} \\[-4pt]
  (Int) & 0.029 & 0.051 & 0.127 & 1.125 & 0.95 & 0.882 \\[-4pt]
  Urban & 0.059 & 0.061 & 0.139 & 1.241 & 0.91 & 0.926 \\[-4pt]
  Healthcare inaccessibility & 0.006 & 0.077 & 0.171 & 1.614 & 0.81 & 0.823 \\[-4pt]
  Elevation & 0.008 & 0.025 & 0.090 & 0.957 & 0.90 & 0.546 \\[-4pt]
  Dist.~rivers \& lakes & -0.014 & 0.015 & 0.069 & 0.595 & 0.93 & 0.467 \\[-4pt]
  Population & -0.037 & 0.150 & 0.224 & 1.832 & 0.90 & 1.343 \\[-4pt]
    \multicolumn{7}{l}{\textbf{M\textsubscript{\tiny DM}}} \\[-4pt]
  (Int) & 0.010 & 0.009 & 0.055 & 0.439 & 0.94 & 0.353 \\[-4pt]
  Urban & 0.050 & 0.012 & 0.062 & 0.487 & 0.91 & 0.364 \\[-4pt]
  Healthcare inaccessibility & 0.002 & 0.004 & 0.035 & 0.285 & 0.95 & 0.230 \\[-4pt]
  Elevation & 0.011 & 0.005 & 0.039 & 0.367 & 0.92 & 0.243 \\[-4pt]
  Dist.~rivers \& lakes & -0.001 & 0.003 & 0.032 & 0.266 & 0.96 & 0.224 \\[-4pt]
  Population & -0.025 & 0.016 & 0.071 & 0.604 & 0.92 & 0.449 \\[-4pt]
    \multicolumn{7}{l}{\textbf{IHME\textsubscript{\tiny SPDE}}} \\[-4pt]
  (Int) & 0.355 & 0.150 & 0.251 & 2.292 & 0.63 & 1.086 \\[-4pt]
  Urban & 0.439 & 0.198 & 0.406 & 13.186 & 0.00 & 0.230 \\[-4pt]
  Healthcare inaccessibility & 0.006 & 0.000 & 0.010 & 0.090 & 1.00 & 0.090 \\[-4pt]
  Elevation & 0.000 & 0.001 & 0.017 & 0.165 & 0.96 & 0.147 \\[-4pt]
  Dist.~rivers \& lakes & -0.006 & 0.001 & 0.022 & 0.163 & 0.95 & 0.147 \\[-4pt]
  Population & -0.433 & 0.189 & 0.409 & 14.158 & 0.00 & 0.168 \\[-4pt]
    \multicolumn{7}{l}{\textbf{IHME\textsubscript{\tiny BYM2}}} \\[-4pt]
  (Int) & 0.402 & 0.163 & 0.358 & 10.166 & 0.00 & 0.311 \\[-4pt]
  Urban & 0.438 & 0.195 & 0.404 & 13.099 & 0.00 & 0.232 \\[-4pt]
  Healthcare inaccessibility & 0.004 & 0.000 & 0.007 & 0.088 & 1.00 & 0.088 \\[-4pt]
  Elevation & -0.001 & 0.005 & 0.000 & 0.009 & 1.00 & 0.107 \\[-4pt]
  Dist.~rivers \& lakes & 0.000 & 0.000 & 0.009 & 0.106 & 1.00 & 0.106 \\[-4pt]
  Population & -0.445 & 0.198 & 0.422 & 14.727 & 0.00 & 0.162 \\[-4pt]
  \bottomrule
\end{tabular}
}
\vspace{-6pt}
\caption{BYM2 simulation study parameter estimate scores and metrics with respect to simulation parameters.}
\label{tab:sim1parBYM2}
\end{table}

\FloatBarrier

\renewcommand{\arraystretch}{1.0}

\section{Additional results for the application}


\begin{table}[ht]
\centering
\resizebox{\linewidth}{!}{
\begin{tabular}{lcccccc|ccc}
\toprule
 & \textbf{(Int)} & \textbf{Urban} & \textbf{Health.~Inaccess.} & \textbf{Elevation} & \textbf{Dist.~Riv.~Lakes} & \textbf{Population} & $\boldsymbol{\sigma^2}$ & $\boldsymbol{\phi}$ & $\boldsymbol{\sigma_{\epsilon}^2}$ \\
\midrule
\textbf{Est} & -1.20 & 1.65 & -0.02 & 0.04 & 0.03 & 0.15 & 0.70 & 0.92 & 1.90 \\
\textbf{95\% CI} & (-1.34, -1.05) & (1.51, 1.79) & (-0.08, 0.04) & (-0.03, 0.11) & (-0.04, 0.09) & (0.04, 0.26) & \multicolumn{1}{c}{--} & \multicolumn{1}{c}{--} & \multicolumn{1}{c}{--} \\
\bottomrule
\end{tabular}
}
\caption{Posterior parameter central estimates and 95\% credible intervals for the IHME\textsubscript{\tiny BYM2} model (runtime 0.60 minutes).}
\label{tab:parSummaryIHMEBYM2}
\end{table}

\FloatBarrier

\FloatBarrier

\section{Additional validation results} \label{sec:addVal}

\FloatBarrier

\bibliographystyle{apalike}

\bibliography{myBib}